\documentclass[sigconf]{acmart}

\usepackage{changepage}
\usepackage{longtable}
\usepackage{soul}
\usepackage[capitalize]{cleveref}
\usepackage[shortlabels]{enumitem}
\setlist[enumerate]{nosep}
\usepackage{subcaption, multirow, amsfonts,dcolumn}
\usepackage{colortbl}
\usepackage{fontawesome5}
\usepackage{verbatim}
\usepackage{xspace}
\usepackage{listings}
\usepackage{framed}
\usepackage{pifont}

\soulregister\ref7
\soulregister\textit7
\soulregister\cite7

\crefname{subsection}{subsection}{subsections}
\Crefname{subsection}{Subsection}{Subsections}

\newif\ifnewchanges
\sethlcolor{yellow}

\AtBeginDocument{%
  }

\copyrightyear{2026}
\acmYear{2026}
\setcopyright{cc}
\setcctype{by}
\acmConference[ICER 2026 Vol. 1]{Proceedings of the ACM Conference on International Computing Education Research Vol.1}{August 11--14, 2026}{Uppsala, Sweden}
\acmBooktitle{Proceedings of the ACM Conference on International Computing Education Research Vol.1 (ICER 2026 Vol. 1), August 11--14, 2026, Uppsala, Sweden}
\acmDOI{10.1145/3765964.3811662}
\acmISBN{979-8-4007-2203-5/2026/08}

\begin{document}

\newcommand{\pvalue}[1]{%
  \ifdim#1pt>0.05pt
    \textit{p} = #1%
  \else
    \ifdim#1pt<0.001pt
      \textit{\textbf{p}} < .001%
    \else
      \textit{\textbf{p}} = #1%
    \fi
  \fi
}

\newcommand{\chiresult}[3]{%
  \(\chi^2\)(#2, \textit{N}=508) = #1, \pvalue{#3}%
}

\newcommand{\olsresult}[6]{%
  Coefficient: \(b=#1\) (\textit{SE}=#2), \(t(#3)=#4\), \pvalue{#5}, \(R^2=#6\)%
}


\newcounter{quotecount}
\newcommand{\quoteinc}{%
  \stepcounter{quotecount}%
  \textbf{Quote \arabic{quotecount}: }%
}

\newenvironment{myquote}
  {\begin{quote}\itshape}
  {\end{quote}}

\makeatletter
\newcommand\quotefontsize{%
  \@setfontsize\quotefont{8.8}{11}%
}
\makeatother


\definecolor{shadecolor}{gray}{0.95}

\definecolor{quotehead}{gray}{0.82}   
\definecolor{quotebody}{gray}{0.96}   

\newenvironment{shadedcolored}[1]{%
  \def\FrameCommand{\fboxsep=6pt\colorbox{#1}}%
  \MakeFramed{\FrameRestore}%
}{\endMakeFramed}

\newenvironment{myquote2}[1][]{%
  \def\quotelabel{#1}%
  \par\vspace{6pt}%
  \noindent
  \begin{shadedcolored}{quotehead}%
    \noindent\small\textbf{\quoteinc\;}\textnormal{\quotelabel}%
  \end{shadedcolored}%
  \vspace{-12pt}
  \begin{shadedcolored}{quotebody}%
    \noindent\normalsize\ignorespaces
}{%
  \end{shadedcolored}%
  \vspace{6pt}%
}
  

\newcommand{\newimage}[3]{%
  \begin{center}
    \includegraphics[width=0.48\textwidth]{#1}
    \captionof{figure}{#2}
    \label{fig:#3}
  \end{center}
}

\newcommand{\newbigimage}[5][1]{%
  \begin{center}
    \scalebox{#1}{%
      \includegraphics[width=0.4\textwidth]{#2}%
    }%
    \captionsetup{hypcap=false}%
    \captionof{figure}{#3}%
    \Description{#4}%
    \label{fig:#5}%
  \end{center}%
}

\newcommand{\newpicture}[3]{%
  \begin{center}
    \includegraphics[width=0.8\textwidth]{#1}
    \captionof{figure}{#2}
    \label{fig:#3}
  \end{center}
}

\newcommand{\mymultirow}[2]{#1 #2}

\long\def\comment#1{}

\sethlcolor{yellow}
\newcommand{\hlpink}[1]{\sethlcolor{pink}\hl{#1}\sethlcolor{yellow}}  
\newcommand{\hlblue}[1]{\sethlcolor{cyan!30}\hl{#1}\sethlcolor{yellow}}
\newcommand{\hlgreen}[1]{\sethlcolor{lime}\hl{#1}\sethlcolor{yellow}}
\newcommand{\hlgray}[1]{\sethlcolor{gray!30}\hl{#1}\sethlcolor{yellow}}

\newcommand{\bibncite}[1]{\bibentry{#1} \cite{#1}}

\definecolor{btnback}{HTML}{D6F5F5}
\definecolor{btnframe}{HTML}{106B7A}

\newcommand{\inlinebutton}[1]{%
  {\setlength{\fboxsep}{2pt}%
   \setlength{\fboxrule}{0.5pt}%
   \raisebox{-0.3ex}{\fcolorbox{btnframe}{btnback}{\strut\,#1\,}}}%
}
\definecolor{softcyan}{HTML}{D9FFFF} 

\newcommand{\circnum}[1]{%
  \raisebox{0.5pt}{%
    \setlength{\fboxsep}{1.5pt}%
    \colorbox{softcyan}{%
      \makebox[1.2em][c]{\bfseries\small #1}%
    }%
  }%
}

\definecolor{codeblue}{HTML}{1F77B4}   
\definecolor{memorygreen}{HTML}{2CA02C} 
\definecolor{metaphorpurple}{HTML}{9467BD} 

\newcommand{\Code}{\textcolor{codeblue}{\texttt{Code}}\xspace}
\newcommand{\Memory}{\textcolor{memorygreen}{\texttt{Memory}}\xspace}
\newcommand{\Metaphor}{\textcolor{metaphorpurple}{\texttt{Metaphor}}\xspace}

\definecolor{sand}{RGB}{243,239,230}

\definecolor{taupe}{RGB}{225,220,215} 

\title[
  Code as Anchor, Memory, and Metaphor as Support: Learner Experiences with Multi-View Visualizations
]{%
  \texorpdfstring{%
    Code as Anchor, Memory and Metaphor as Support: \\
    Learner Experiences with Multi-View Visualizations%
  }{%
    Code as Anchor, Memory and Metaphor as Support: Learner Experiences with Multi-View Visualizations%
  }%
}
\author{Naaz Sibia}
\orcid{0000-0001-7628-7077}
\affiliation{
  \institution{University of Toronto}
  \city{Toronto}
  \country{Canada}
}
\email{naaz.sibia@utoronto.ca}

\author{Jessica Wen} 
\orcid{0009-0000-2766-5916}
\affiliation{
  \institution{University of Toronto}
  \city{Mississauga}
  \country{Canada}
}
\email{jessica.wen@mail.utoronto.ca}

\author{Amber Richardson}
\orcid{0009-0003-9431-9126}
\affiliation{
  \institution{University of Toronto}
  \city{Mississauga}
  \country{Canada}
}
\email{amber.richardson@mail.utoronto.ca}

\author{Yashika Jain} 
\orcid{0009-0001-6787-3960}
\affiliation{
  \institution{University of Toronto}
  \city{Mississauga}
  \country{Canada}
}
\email{yashika.jain@mail.utoronto.ca}

\author{Khushi Malik}
\orcid{0009-0008-5336-5819}
\affiliation{
\institution{University of Toronto}
   \city{Mississauga}
   \country{Canada}
}
\email{khushi.malik@mail.utoronto.ca}

\author{Bogdan Simion}
\orcid{0000-0002-2554-8705}
\affiliation{%
 \institution{University of Toronto}
 \city{Mississauga}
 \country{Canada}
}
\email{bogdan@cs.toronto.edu}

\author{Carolina Nobre}
\orcid{0000-0002-2892-0509}
\affiliation{
  \institution{University of Toronto}
  \city{Toronto}
  \country{Canada}
}
\email{carolina.nobre@utoronto.ca}

\author{Angela Zavaleta Bernuy}
\orcid{0000-0002-1228-5774}
\affiliation{
  \institution{McMaster University}
  \city{Hamilton}
  \country{Canada}
}
\email{zavaleta@mcmaster.ca}

\author{Andrew Petersen}
\orcid{0000-0003-1337-7985}
\affiliation{
  \institution{University of Toronto}
  \city{Mississauga}
  \country{Canada}
}
\email{andrew.petersen@utoronto.ca}

\author{Michael Liut}
\orcid{0000-0003-2965-5302}
\affiliation{
  \institution{University of Toronto}
  \city{Mississauga}
  \country{Canada}
}
\email{michael.liut@utoronto.ca}

\renewcommand{\shortauthors}{Sibia, Wen, Richardson, Jain, Simion, Nobre, Zavaleta Bernuy, Petersen, and Liut}




\begin{abstract}
\textbf{Motivation:} Program visualizations are widely used to support novice programmers, yet students often ignore or resist well-designed visual scaffolds. Research on multiple external representations (MERs) suggests cognitive design principles for coordinating views, but says little about what determines whether learners actually engage with the representations available to them. \\ 
\textbf{Method:} We conducted a within-subjects study with 19 undergraduates who had completed CS1 and CS2, combining think-aloud tasks, reflective interviews, and webcam-based gaze tracking as students worked with a multi-representational probe (synchronized code, memory, and metaphor views) and Python Tutor across three topics (scope, while loops, and linked lists). \\ 
\textbf{Findings:} Gaze analysis showed students spent nearly half their time focused on code despite available visual scaffolds, with students without prior experience anchoring even more heavily in code and engaging minimally with metaphor views. Interview accounts revealed three themes explaining this selective engagement: students sought control over their own cognitive effort rather than having it reduced (agency), responded to identical designs with wide variation in what felt helpful versus overwhelming (representational fit), and avoided metaphorical scaffolds they perceived as childish or insufficiently rigorous for university-level work (legitimacy). \\ 
\textbf{Implications:} These findings suggest that deploying multi-representational tools in computing education requires attention to affective and social factors alongside cognitive design. Practical considerations include positioning visualizations as verification instruments rather than explanation guides, providing toggleable abstraction levels so students can calibrate complexity to their needs, and framing tools to signal disciplinary legitimacy. More broadly, our themes offer a starting point for investigating why cognitively sound visualization tools sometimes fail to engage the students they are designed to help.
\end{abstract}

\maketitle

\section{Introduction}
\label{sec:intro}

Learning to program can require reasoning across multiple levels of abstraction, such as simulating execution step by step, predicting behavior, and explaining outcomes in terms of broader program purpose \cite{hayatpur2023crosscode, johnson1983mental, seel2006mental}. Expert developers shift flexibly between these abstraction levels \cite{yates2020characterizing, sillito2008asking, von1993code} and sketch visual metaphors to support that shift \cite{ma2020domain}, but novices often struggle to move between concrete traces and high-level conceptual models \cite{levy2019understanding, sajaniemi2008study}. This difficulty is well documented in core CS1 and CS2 topics such as scope, iteration, and linked data structures \cite{cherenkova2014identifying, morazan2020make, zingaro2018identifying}, where students must coordinate reasoning about program purpose and runtime state. 

Visualization tools have long been proposed to help bridge this gap. Widely used tools like Python Tutor \cite{guo2013online} make execution state visible, but students still report confusion with visual encodings such as arrows and call stacks \cite{karnalim2017effectiveness, karnalim2017use}, and such tools show \textit{how} a program behaves in memory without helping students see \textit{why} it behaves that way or \textit{what} its purpose is \cite{pollock2020essence, sorva2013review}. Research in the learning sciences suggests that coordinating multiple representations (linking concrete execution to more abstract depictions) can aid understanding and reduce cognitive load \cite{ainsworth2006deft, card1999readings, mnguni2014theoretical}. The theory of \textit{Multiple External Representations} (MERS) and its DeFT framework \cite{ainsworth2006deft} formalize this idea, and MER-based approaches have shown benefits in mathematics \cite{tripathi2008developing}, physics \cite{treagust2017multiple}, and chemistry \cite{permatasari2022chemistry}, and our recent deployments in introductory programming courses suggest similar potential for coordinated code, memory, and analogy views~\cite{sibia2025code, sibia2025state}.

Yet a persistent gap remains: students often ignore or resist well-designed visualizations \cite{schwonke2009multiple}, even when those visualizations are designed using cognitive design principles. MER frameworks focus on how representations support learning \textit{given engagement}, but they say little about what determines whether learners actually engage. In formal educational settings, where students are still developing their disciplinary identities, they may not engage with available support. Prior work has shown that tool engagement in software development reflects psychological needs like autonomy and competence~\cite{wong2025spectrum}, and that multi-level visualizations benefit experienced programmers~\cite{hayatpur2023crosscode}, but how these factors operate for students in introductory programming courses remains underexplored.

We address this gap by investigating the factors that shape whether and how CS1/CS2 students engage with multi-view program visualizations, asking:

\begin{description}[leftmargin=0.2cm, labelsep=0.2cm]
\item[\textbf{RQ1:}] How do students distribute their attention across code, memory, and metaphor views when working with a multi-representational visualization tool?
\item[\textbf{RQ2:}] What factors shape whether and how students engage with or resist different representational forms?
\end{description}

To investigate these questions, we designed and conducted our study in two stages. First, we developed a multi-representational design tool, refined through formative feedback from instructors and teaching assistants, that synchronizes three views: code, memory state, and visual analogy. Second, we conducted within-subject sessions with 19 undergraduates who had recently completed CS1 and CS2 at a
large North American research university. Each session combined task-based problem solving, think-aloud protocols, reflective interviews, and webcam-based gaze tracking as students engaged with both our tool and Python Tutor. We take an interpretive stance; our goal is not to identify the ``best'' tool, but to understand how students make sense of visualizations in relation to their prior experiences, disciplinary identities, and the problems at hand.
\section{Related Work}
\label{sec:related}

We draw on work from the learning sciences (multiple external representations, cognitive load), computing education (mental models, notional machines, novice and expert differences), and program visualization research. Together, these strands motivate our study by providing theoretical foundations for how multiple views can support learning, highlighting challenges novices face when forming program models, and situating the design space our probe occupies.

\subsection{Theoretical Foundations}

Visual representations are powerful cognitive supports. They can offload effort and memory demands~\cite{scaife1996external}, support pattern recognition~\cite{munzner2014visualization, ware2012information}, and promote deeper conceptual understanding~\cite{heer2012interactive, mnguni2014theoretical}. In Mnguni's \textit{cognitive process of visualization}, learners internalize visuals, connect them to prior knowledge, and externalize mental models via sketches, highlighting the active engagement that visualization can foster~\cite{mnguni2014theoretical}.

While a single view can aid learning, research in mathematics, physics, and chemistry shows that \textit{multiple coordinated views} yield greater benefits, supporting understanding of abstract concepts such as functions, forces, and molecular structures~\cite{tripathi2008developing, treagust2017multiple, permatasari2022chemistry}. \textbf{Multiple External Representations} (MERs) theory formalizes these findings, explaining how learners coordinate complementary but distinct sources of information~\cite{ainsworth_functions_1999, ainsworth2006deft}. 
Ainsworth's DeFT framework identifies three functions that multiple representations serve. \textit{Complementary} representations show different representations. For instance, a memory diagram and a call stack each reveal aspects of execution that the other does not. \textit{Constraining} representations help learners interpret an unfamiliar view by pairing it with a familiar one, so a student who understands code can use it to make sense of an abstract diagram they have not seen before. \textit{Constructive} use happens when the act of translating between representations itself produces understanding, not just reading two views, but connecting them. The design dimension of the framework stresses reducing redundancy, guiding attention, and preventing overload from excessive or misaligned views~\cite{mayer2008increased, wu2012affordances}, while the tasks dimension underscores the demanding but essential work of interpreting, relating, and integrating views into coherent mental models. Because of this dual focus on design and learner activity, DeFT has become a central lens for evaluating MER-based tools, including those developed for programming education~\cite{suh2022codetoon, suh2020coding}.

In computing education specifically, MER-based deployments have shown similar benefits, with coordinated code, memory, and analogy views reducing immediate mental effort and supporting engagement in introductory programming courses~\cite{sibia2025code, sibia2025state}.

\textbf{Cognitive Load Theory} (CLT) complements DeFT by distinguishing  three kinds of mental effort~\cite{paas2014cognitive, sweller2011cognitive}.  \textit{Intrinsic load} refers to the complexity of the material itself. \textit{Extraneous load} is the unnecessary effort caused by poor design. \textit{Germane load} was originally proposed as a third, independent component representing productive effort that supports learning, but contemporary CLT has revised this: germane load is better understood as the portion of working memory directed toward managing intrinsic load, not a separate resource on top of it~\cite{duran2022cognitive}. The practical implication across all formulations remains the same: minimize extraneous demands so learners can focus on the material. Learner expertise strongly shapes these effects. Supports that help novices can actively hinder experts, a phenomenon known as the expertise reversal effect~\cite{kalyuga2007expertise}, which means visualizations need to be calibrated to who is using them, not just what they are showing~\cite{adolph2015gibson, mnguni2014theoretical}.

Taken together, MER theory, DeFT, and CLT point toward a common design principle: effective programming visualizations should provide enough complementary views to support abstraction but not so many that they overwhelm, reduce redundancy and guide attention with clear cues, sequence from concrete to abstract, and tailor support to a learner's level of expertise~\cite{baldonado2000guidelines, fyfe2014concreteness, hassan2024evaluating}. These frameworks provide strong guidance for the \textit{cognitive} design of multi-representational tools. However, they say relatively little about the \textit{affective} and \textit{social} dimensions of engagement. Recent work by \citet{wong2025spectrum} has shown that tool engagement in software development contexts reflects psychological needs such as autonomy and competence, suggesting that uptake involves more than cognitive fit. 

\subsection{Mental Models in Programming}

Programmers require layered mental models that link low-level execution (e.g., memory updates) with broader functional and structural goals~\cite{johnson1983mental, brooks1978using, canas2001role, seel2017model}. Such models include program models (runtime behavior), concept models (abstract ideas such as recursion), and task or solution models (strategies for goals)~\cite{heinonen2023synthesizing}. Schema-based perspectives frame them as structured templates that help learners interpret and anticipate behavior~\cite{detienne1990program, balijepally2012effect}. Novices often fall back on surface cues or line-by-line tracing~\cite{levy2019understanding}, whereas experts integrate local and global reasoning flexibly~\cite{von1993code, sillito2008asking, yates2020characterizing}.

In CS1~\cite{hertz2010cs1}, mental models are scaffolded by \emph{notional machines}, idealized representations of language behavior that support reasoning about scope, control flow, and function calls~\cite{juha2013notional, fincher2020notional}. Although partial and idiosyncratic~\cite{greca2000mental, eckerdal2005novice}, notional machines provide a foundation for deeper understanding. Without such scaffolds, detailed execution traces (e.g., those produced by Python Tutor~\cite{guo2013online}) risk focusing attention on low-level mechanics rather than conceptual ideas~\cite{karnalim2017effectiveness, pollock2020essence, eckert2022loops}.

In CS2~\cite{porter2018developing}, recursion, references, and hierarchical data structures require students to connect local operations to global structural intent~\cite{zingaro2018identifying, almadhoun2021exploratory}. MERs can help by scaffolding \emph{representational competence} (the ability to interpret and construct representations) and \emph{representational fluency} (the ability to translate flexibly among them)~\cite{ainsworth2006deft, daniel2018representational, zaqoot2019representational, schwonke2009multiple}. However, building such fluency is itself cognitively demanding, and students do not always engage with the representations available to them~\cite{schwonke2009multiple}. 

Isohanni and Knobelsdorf's qualitative study of the VIP visualization tool offers a unique long-term perspective on this phenomenon~\cite{isohanni2013students}. Tracking students across a full semester, they identified four phases of tool engagement (introductory, progressive, creative, and routine use) and eight working patterns ranging from deep dynamic debugging to superficial output-checking. Interpreted through Activity Theory, their central finding is that a visualization tool functions as a useful mediator primarily during the internalization process. Students who automatized low-level tool interactions and developed personal uses of the tool benefited most, while those who never moved beyond instructor-prescribed routines did not. A subsequent survey of over 250 programming teachers confirmed the broader adoption problem: roughly 20\% of courses used visualization tools regularly, and the most common barrier was not technical but psychological, as teachers preferred their own externalizations of programming concepts over those embedded in third-party tools~\cite{isohanni_are_2014, knobelsdorf2012reasons}. Together, these studies underscore that engagement is shaped not only by tool design but by long-term learning dynamics and individual agency, factors our study investigates in a multi-representational context.

\subsection{Program Visualization Tools}
\label{existing-tools}

Interactive program visualizations have a long history in computing education, with early systems like BALSA~\cite{brown1984system} establishing the paradigm of program-generated animations for teaching fundamental concepts. This tradition produced a rich lineage of systems offering multiple coordinated views and varying levels of abstraction, including HalVis~\cite{hansen2002designing}, which combined conceptual, detailed, and populated views with interactive prediction; SRec~\cite{velazquez2008srec}, which coordinated three views of recursion with typographic linking hints drawn from information visualization; and later web-based systems like JHAVÉ, TRAKLA2, and Animal that emphasized interactive exercises and feedback~\cite{shaffer2010algorithm}. Early work pairing animations with explanations improved novices' debugging success~\cite{brusilovsky1993program}, demonstrating the benefit of linking visuals with text. Sirkiä's Jsvee and Kelmu libraries extended this tradition to the web, offering expression-level animations augmented with instructor-authored annotations; empirical studies found that annotations increased time-on-task at key steps and were read by over 80\% of students, though observations of learning gains were statistically inconclusive~\cite{sirkia2018jsvee}. A recent large-scale field deployment of an MER-based tool in CS1 found that synchronized code, memory, and analogy views reduced immediate mental effort compared to text-only conditions and yielded higher engagement than both single-view and text-only approaches, with benefits moderated by language background and prior experience~\cite{sibia2025code}.

A central lesson from this body of work, consolidated by Hundhausen, Douglas, and Stasko's meta-study of 24 experiments~\cite{hundhausen2002meta}, is that \textit{how} students use a visualization matters more than \textit{what} it shows them. Experiments manipulating representational features produced significant learning effects far less reliably than those manipulating learner activity. Building on this, Naps et al.~\cite{naps2002exploring} proposed an engagement taxonomy, arguing that visualization tools ``are of little educational value unless they engage learners in an active learning activity.'' Subsequent reviews~\cite{shaffer2010algorithm} found that although many algorithm visualizations exist, coverage is uneven and the pedagogical lessons of this literature are not always propagated to new tools. Beyond engagement design, learner-side factors may also shape uptake. Clancy's survey of programming misconceptions and attitudes \cite{clancy2005misconceptions} documents how novices' tendency to \textit{avoid} rather than \textit{manage} complexity, along with beliefs that richer structures like arrays of records make programs harder to read, can lead students to reject the abstractions that we assume will help them.

More recent systems embed runtime state directly into the code: Omnicode shows variable values inline~\cite{kang2017omnicode}, and Projection Boxes extend this with customizable views~\cite{lerner2020projection}. These tools improve debugging efficiency but offer limited support for cross-representational reasoning or abstraction. At a different level, Ko and Myers' Whyline reframes debugging as asking questions~\cite{ko2009whyline}, and Ma'ayan et al.~\cite{ma2020domain} study how experts sketch and adapt diagrams, proposing a ``natural diagramming'' framework that connects to calls in computing education for making notional machines more explicit~\cite{sorva2013review}.

Eye-tracking studies highlight why design choices matter for engagement. Beginners often focus heavily on code and neglect diagrams unless explicitly directed to them~\cite{schwonke2009multiple, bednarik2012expertise}. Layout shapes attention: less linear arrangements trigger vertical scanning and extra effort, while experienced learners follow execution-driven strategies~\cite{bansal2021eye, bednarik2012expertise}. More broadly, eye gaze has been shown to reflect cognitive processing, workload, and comprehension~\cite{just1976eye, underwood2004inspecting, ahn2020towards}; in programming contexts specifically, longer fixation durations and shorter saccade amplitudes (the angular distance the eye travels in a rapid movement) have been linked to higher intrinsic load during code comprehension~\mbox{\cite{andrzejewska2020examining}}, suggesting that gaze captures not just where attention lands but how much the material demands. Most directly relevant to our work, CrossCode~\cite{hayatpur2023crosscode} demonstrated that multi-level visualizations help experienced programmers navigate abstraction layers. \citet{hegarty2010thinking} found that visual highlighting only improved performance once participants already understood the underlying concepts. This suggests that representational design choices in programming visualizations may similarly depend on learners having sufficient mental models to use them.

\textit{Positioning}: Across this literature, a consistent gap emerges. MER theory and CLT provide strong guidance for the cognitive design of multi-representational  tools, and mental model research highlights the need to connect execution with conceptual structure. But existing program visualization tools largely emphasize tracing and debugging, and the frameworks that guide their design say relatively little about why students engage with or resist the representational supports available to them.  \citet{knobelsdorf2012reasons} hint at this: even well-designed tools go unused when students never move beyond prescribed routines, and teachers resist tools that don't match their own externalizations. These are not cognitive problems but affective and social ones. Understanding what mediates engagement in multi-representational programming tools, beyond design alone, is the central concern of our study.

\section{Methods}
\label{sec:methods}

We conducted a think-aloud study to examine how students engaged with a multi-representational visualization tool alongside the widely used Python Tutor visualizer. Following a \textit{within-subjects design}, all participants used both tools, allowing us to compare their experiences directly. We focused on documenting learners' reasoning processes: where they paid attention, where they got stuck, and how they reacted across representations.

\subsection{Visualization Tools}
\label{subsec:tools}

Participants worked with two tools: a multi-representational visualization tool designed for this study, and Python Tutor as a comparison baseline.

\subsubsection{Multi-Representational Tool (MR)}

We designed a visualization tool with multiple views\footnote{All visualization materials are openly available at \href{github.com/CORE-Research-Lab/Interactive-Viz-IntroProgramming}{https://github.com/CORE-Research-Lab/Interactive-Viz-IntroProgramming}.}
~that draws on MER/DeFT~\cite{ainsworth2006deft} and concreteness fading~\cite{fyfe2014concreteness} to address limitations of existing program visualization tools, which often support tracing but provide little scaffolding for cross-representational reasoning. As shown in Figure~\ref{fig:design-feedback}, the tool was refined across three formative feedback cycles with five instructors, a visualization researcher, and nine teaching assistants from CS1 and CS2, who reviewed prototypes for conceptual fidelity\footnote{We use \emph{fidelity} as a shorthand for how closely a visualization preserves relationships and behaviors considered essential to program execution (e.g., stack frames for function calls, pointer diagrams for references). Fidelity here does not imply literal correspondence to how memory is implemented in hardware, but rather alignment with canonical representations in programming visualizations.}, usability, and alignment with common student misconceptions. Feedback informed refinements such as progressive disclosure \cite{springer2019progressive}, clearer metaphor labels, and stepwise tracing pacing.

\newbigimage[1.2]{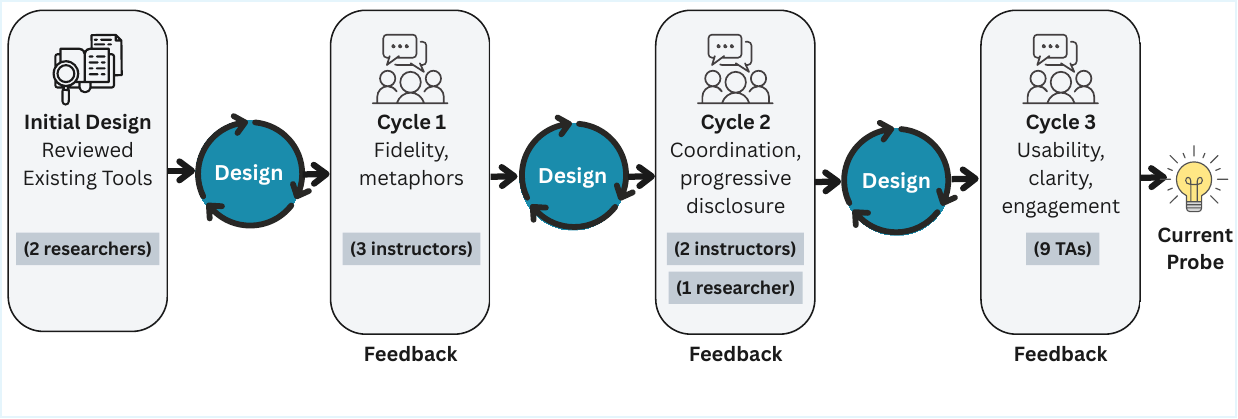}{Iterative design cycles for the multi-representational tool. Each cycle involved different stakeholders (instructors, researcher, TAs) and contributed  feedback that guided refinements to conceptual fidelity, representation coordination, and usability, leading to the tool used in our study.}{A flow diagram showing three iterative design cycles leading to the final multi-representational tool. The diagram begins with ``Initial Design,'' informed by a review of existing tools. Three feedback cycles follow: Cycle 1 (fidelity and metaphors), Cycle 2 (coordination and progressive disclosure), and Cycle 3 (usability, clarity, and engagement). Each cycle includes feedback from different stakeholders, including instructors, a visualization researcher, and teaching assistants. Arrows indicate refinement across cycles, culminating in the ``Current Tool.''}{design-feedback}

\newbigimage{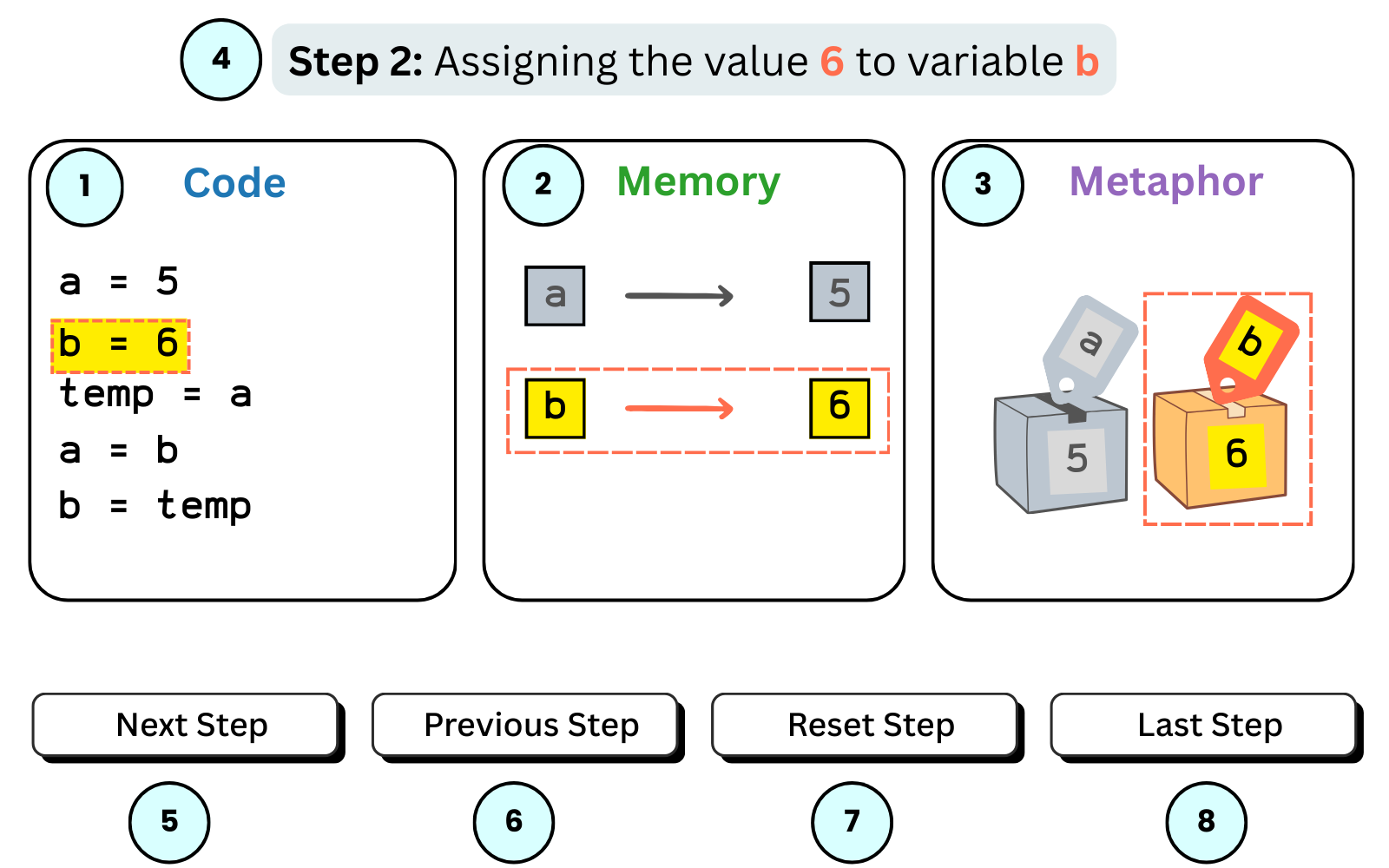}{The multi-representational visualization tool, showing three synchronized panels: (1) Code View, (2) Memory View, and (3) Metaphor View, alongside interactive controls (5--8).}{Screenshot of the multi-representational visualization tool interface. The screen is divided into three synchronized panes: (1) Code View showing Python source code with the active line highlighted in yellow with an orange outline; (2) Memory View displaying variables, values, and object references similar to Python Tutor; and (3) Metaphor View showing an analogy-based visual representation of program state. Below or alongside the panes are execution controls labeled Next Step, Previous Step, Reset, and Last Step. The panes are coordinated so that selecting elements in one highlights corresponding elements in the others.}{tool-layout}

The interface (Figure~\ref{fig:tool-layout}) presents three synchronized panes:

\circnum{1} \textbf{Code View} displays Python source code with the active line highlighted (yellow fill, orange outline) to guide attention and support traceability.

\circnum{2} \textbf{Memory View} depicts variable bindings and object references in the program's current state, closely aligned with Python Tutor's representation to preserve familiarity.

\circnum{3} \textbf{Metaphor View} uses analogy-based depictions of program state, drawing on metaphors from prior work~\cite{sanford2014metaphors, hermans2018thinking}. For scope management, a ``house'' metaphor depicts each scope as a separate room, with function calls shown as animated arrows between rooms. For while loop iteration, a ``magician's table'' scene reveals list elements in sequence until the target is found. For linked lists, nodes are shown as connected boxes with arrows reflecting pointer references, with a relay race metaphor to make the sequential handoff between nodes more salient. These metaphors were selected through formative feedback to balance engagement with conceptual fidelity, and were chosen for broad recognizability across diverse student backgrounds.

Execution is synchronized across all three views and controlled through four buttons: \texttt{Next Step} advances execution incrementally (line-level for CS1, function-level for CS2); \texttt{Previous Step} reverses by one step; \texttt{Reset} returns to the initial state; and \texttt{Last Step} executes the remainder at once. Hover and selection events are linked across views (brushing and linking~\cite{roberts2006towards}), so that selecting a variable in the Code View highlights its memory cell and the corresponding metaphor element. This coordination follows a longer tradition of applying InfoVis interaction techniques to program animation, where coordinated views are linked through shared visual conventions and synchronized state to support cross-representational reasoning \cite{velazquez2010infovis}.

\subsubsection{Comparison Baseline.}

\begin{figure}[h]
    \centering
    \includegraphics[width=0.85\columnwidth]{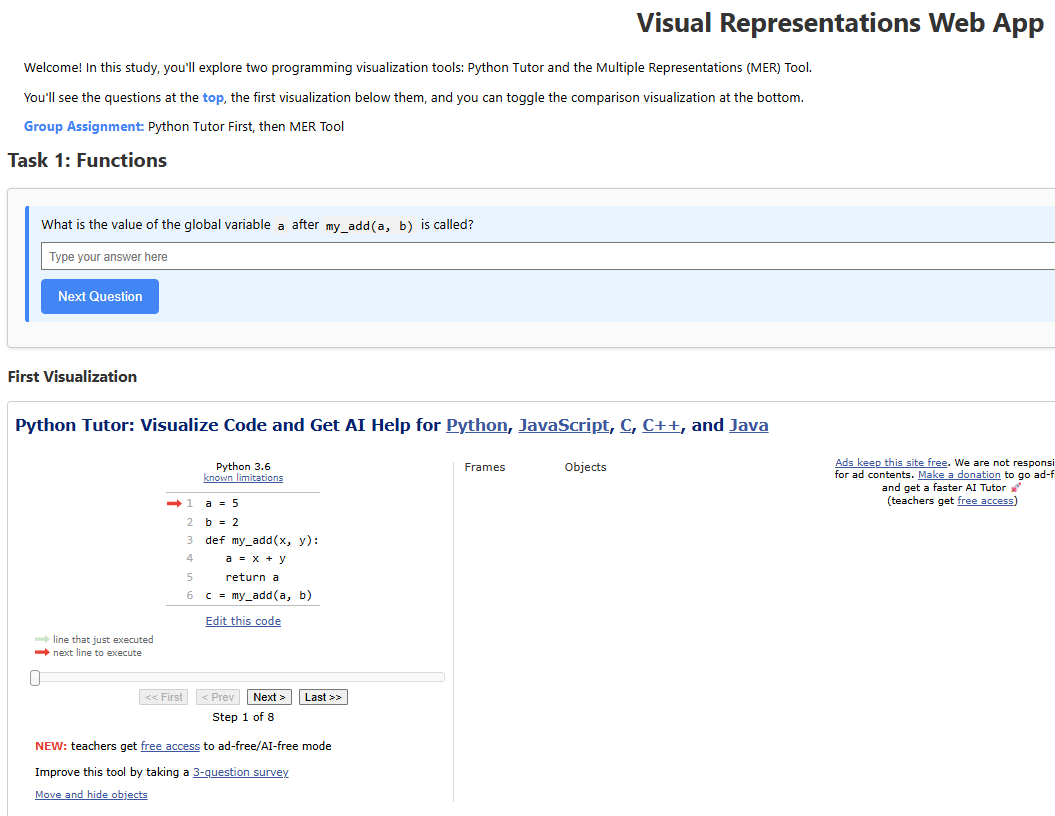}
    \caption{Python Tutor visualization as seen by participants,
showing line-by-line execution with variable state and stack frames.}
    \Description{Screenshot of the Python Tutor interface used as a comparison baseline. The left side shows Python source code with the current line highlighted. The right side displays a memory diagram with stack frames and variables, including arrows indicating object references. Execution controls allow stepping forward and backward line by line. The interface emphasizes precise runtime state but does not include additional abstraction layers or metaphorical views.}
    \label{fig:viz-pt}
\end{figure}

Python Tutor~\cite{guo2013online}, a widely used single-view tool that shows execution with line-by-line fidelity served as a contrast case  (Figure~\ref{fig:viz-pt}). Its strength is execution detail, but it lacks abstraction layers, leaving learners to integrate patterns across steps on their own~\cite{pollock2020essence}. This contrast surfaces different design values (precision versus abstraction support) and lets us examine how multi-representational coordination shapes sensemaking.

\subsection{Participants}
\label{subsec:participants}

Nineteen students participated in the study, recruited from two campuses of a large North American research university (R1). All participants had completed both CS1 (introduction to programming) and CS2 (introduction to data structures, algorithms, and object-oriented programming) prior to the study in the first and second semesters of the 2024-2025 academic year. Both courses were taught in Python (version 3.10). CS1 covered programming basics such as variables, loops, lists, dictionaries, and basic sorting algorithms; CS2 covered object-oriented programming basics, introduction to recursion, and abstract data structures (e.g., linked lists, n-ary trees, and binary search trees).

Thirteen participants reported having prior programming experience before university (typically a Java course in grade 12 or basics of Python in grade 11), while six reported no prior experience. The sample included six women and thirteen men. Ten participants completed the ABA tool-ordering sequence and nine completed BAB. Table~\ref{tab:participants} provides a detailed breakdown. The tool ordering sequence is explained below.

\setlength{\tabcolsep}{3pt}

\begin{table}[ht]
\centering
\caption{Participant characteristics by campus, gender, prior
experience (PE), and condition order. Campus codes: A = larger campus,
B = smaller campus.}
\label{tab:participants}
\small
\renewcommand{\arraystretch}{1.15}
\rowcolors{2}{gray!8}{}
\begin{tabular}{@{} l c c c c @{\hskip 0.5em} l c c c c @{}}
\toprule
\textbf{ID} & \textbf{Campus} & \textbf{Gender} & \textbf{PE} & \textbf{Order} &
\textbf{ID} & \textbf{Campus} & \textbf{Gender} & \textbf{PE} & \textbf{Order} \\
\midrule
P1  & B & Man   & Y & ABA & P11 & A & Man   & Y & ABA \\
P2  & B & Man   & Y & BAB & P12 & A & Man   & Y & BAB \\
P3  & B & Man   & Y & ABA & P13 & A & Woman & Y & ABA \\
P4  & B & Man   & Y & BAB & P14 & A & Woman & N & BAB \\
P5  & B & Woman & Y & ABA & P15 & A & Woman & N & ABA \\
P6  & A & Man   & Y & BAB & P16 & A & Man   & N & BAB \\
P7  & A & Woman & Y & ABA & P17 & A & Man   & N & ABA \\
P8  & A & Man   & Y & BAB & P18 & A & Man   & N & BAB \\
P9  & A & Man   & Y & ABA & P19 & A & Woman & N & ABA \\
P10 & A & Man   & Y & BAB &     &   &       &   &     \\
\bottomrule
\end{tabular}
\end{table}

\subsection{Procedure}
\label{subsec:proc}

Recruitment emails were distributed by the interviewer to all students who completed CS1 and CS2 (at both campuses) near the end of the CS2 term, with permission from course instructors. Participants scheduled sessions directly and received compensation equivalent to one hour of research participation, consistent with standard university rates. The interviewer had no prior instructional or evaluative relationship with any participant. Sessions were conducted online via Zoom screen sharing to accommodate participants across two university campuses and facilitate broader recruitment. This study was approved by the university's research ethics board (protocol \#00047002).

At the beginning of each session, the interviewer introduced the study, obtained informed consent, and briefly introduced the think-aloud protocol. Participants were encouraged to verbalize their reasoning, uncertainties, and interpretations while working.

\begin{figure*}[t]
  \centering
  \includegraphics[width=0.7\textwidth]{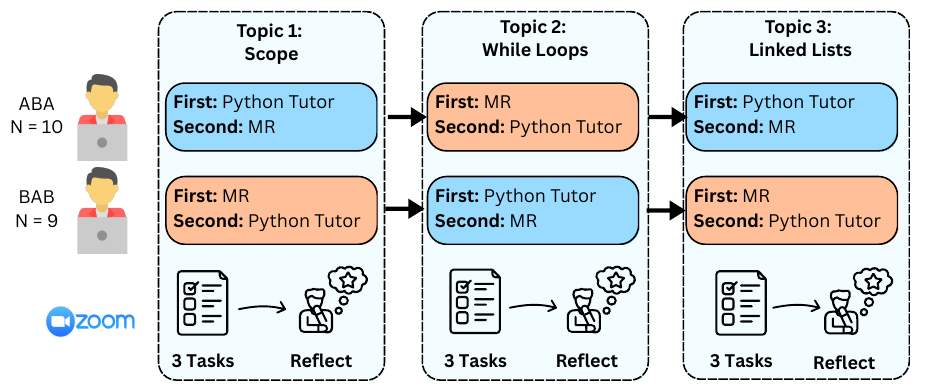}
  \caption{Within-subjects, counterbalanced study design. Participants
(N=19) completed three topical blocks (Scope, While Loops, Linked Lists). Each block comprised three short tasks followed by a brief reflection. Tool order was counterbalanced using ABA and BAB sequences. Sessions were conducted via Zoom screen sharing.}
  \Description{Diagram of the within-subjects counterbalanced study design. Nineteen participants completed three topical blocks in fixed order: Scope, While Loops, and Linked Lists. Within each block, participants completed three tasks followed by a reflection. Tool order was counterbalanced using ABA (Python Tutor, MR, Python Tutor) and BAB (MR, Python Tutor, MR) sequences, with 10 participants in ABA and 9 in BAB. Arrows illustrate progression through topics and alternating tool order.}
  \label{fig:procedure}
\end{figure*}

Each participant worked through three topical blocks in fixed order, chosen to represent progressively more advanced concepts in the CS1 to CS2 sequence: \textbf{Scope} (CS1; tracing variable updates across function calls)~\cite{aedo2016teaching, gupta2023improving}, \textbf{While Loops} (late CS1; reasoning about iteration)~\cite{morazan2020make, cherenkova2014identifying}, and \textbf{Linked Lists} (CS2; traversing pointer-based structures)~\cite{zingaro2018identifying, almadhoun2021exploratory}. These topics are well-documented areas of novice difficulty and require learners to coordinate reasoning about program purpose and runtime state, making them relevant for examining representational integration. Within each block, participants used both tools in counterbalanced order (ABA or BAB sequences), balancing exposure while keeping sessions manageable ($\approx$60 minutes total, with approximately 20 minutes per topic including tasks and reflection). Figure~\ref{fig:procedure} shows the study design.

\subsubsection{Interview Protocol.}
\label{subsec:protocol}

The interviewer, a graduate student with prior experience conducting think-aloud studies in programming courses, used a semi-structured guide to elicit think-alouds and reflections while avoiding overly directive prompts. Prompts targeted \emph{perceptual uptake} (e.g., ``What do you notice in this visual representation?'') and \emph{conceptual integration} (e.g., ``How is the change in variables or control flow depicted for you?''). After completing a task with one tool, participants reflected on its clarity, limitations, and alignment with their reasoning strategies before switching to the other tool. Post-task interviews encouraged cross-tool comparisons (e.g., ``Which tool provided clearer visual cues for you, and why?'') to surface interpretive and affective contrasts. Sessions concluded with an open-ended debrief on perceived usefulness and suggestions for improvement. Table~\ref{tab:protocol} summarizes the key interview prompts used across phases.

\begin{table}[t]
\centering
\caption{Summary of interview prompts by phase. The full protocol is
available in the supplementary repository.}
\label{tab:protocol}
\small
\renewcommand{\arraystretch}{1.3}
\rowcolors{2}{gray!8}{}
\begin{tabular}{@{} p{2.6cm} p{5.4cm} @{}}
\toprule
\textbf{Phase} & \textbf{Example Prompts} \\
\midrule
Perceptual uptake
  & \emph{``What do you notice in this visual representation?''}
    \newline \emph{``What stands out to you first?''} \\
Conceptual integration
  & \emph{``How is the change in variables depicted for you?''}
    \newline \emph{``Can you walk me through what happened in this step?''} \\
Cross-tool comparison
  & \emph{``Which tool provided clearer visual cues, and why?''}
    \newline \emph{``Was there anything one tool showed that the other didn't?''} \\
Debrief
  & \emph{``If you could change one thing about either tool, what would it be?''}
    \newline \emph{``Would you use either of these when studying?''} \\
\bottomrule
\end{tabular}
\end{table}

\subsubsection{Task Design.}
\label{subsec:tasks}

For each topic, task probes asked students to trace, explain, and extend code, balancing execution details with higher-level reasoning. Table~\ref{tab:tasks} shows representative tasks for each topical block.

\begin{table*}[t]
\centering
\caption{Task probes by topic. Each block included three tasks
progressing from tracing to explanation to extension. Tasks were
presented one at a time; participants answered each before proceeding
to the next.}
\label{tab:tasks}
\small
\renewcommand{\arraystretch}{1.15}
\begin{tabular}{@{} >{\bfseries\raggedright\arraybackslash}p{1.8cm}
                    >{\raggedright\arraybackslash}p{1.2cm}
                    p{11cm} @{}}
\toprule
\textnormal{\textbf{Topic}} & \textbf{Type} & \textbf{Task} \\
\midrule
\rowcolor{gray!8}
Scope & Trace & What is the value of the global variable \texttt{a} after
    \texttt{my\_add(a, b)} is called? \\
\rowcolor{gray!8}
      & Trace & What is the value of the local variable \texttt{a} right
    before the return statement? \\
\rowcolor{gray!8}
      & Explain & Explain why the global variable remains unchanged. \\
\addlinespace[1pt]
While Loops & Trace & What value will be printed? Trace each step of the
    loop. \\
            & Explain & Explain in plain English how the while loop searches
    for the item. What role does the counter play, and how does the
    loop stop? \\
            & Extend & How would you modify the code so that if the wanted item
    is not in the list, the program prints `Item not found' instead
    of running indefinitely? \\
\addlinespace[1pt]
\rowcolor{gray!8}
Linked Lists & Trace & What are the values of \texttt{data} and \texttt{next}
    for each node in the list? List them in order. \\
\rowcolor{gray!8}
             & Explain & Explain how the \texttt{append} function works when
    adding a new node, differentiating between an empty list and a
    populated list. \\
\rowcolor{gray!8}
             & Extend & How would you implement a \texttt{find(data)} method,
    and what would happen if the value does not exist? \\
\bottomrule
\end{tabular}
\end{table*}

All sessions were screen-recorded to capture verbalizations, on-screen interactions, and tool states. Audio was transcribed verbatim and anonymized. In addition to screen recordings, we collected gaze data (described in~\Cref{subsec:gaze}) to understand how learners visually engaged with each tool.

\subsection{Analytic Approach}

We used a mixed-method approach. Qualitative data from participants' think-alouds and reflections served as the primary evidence base, analyzed through reflexive thematic analysis. Gaze data were treated descriptively, providing broad markers of visual attention that complemented and triangulated participants' verbal accounts. This design allowed us to connect what students said with how they visually engaged with representations.

\subsection{Gaze Analysis}
\label{subsec:gaze}

We used WebGazer.js~\cite{papoutsaki2018eye} to enable webcam-based eye tracking during online sessions. WebGazer infers gaze position by mapping pupil and head movements to on-screen coordinates through regression-based models. Although less precise than lab-based infrared systems, webcam tracking achieves sufficient accuracy (1--2 cm median error) for identifying broad areas of interest in web-based studies~\cite{papoutsaki2017searchgazer, slim2023moving}, which was appropriate for our design since our areas of interest were the three large visualization panes rather than fine-grained interface elements.

We defined three areas of interest (AOIs) corresponding to the visualization panes of the MR tool: \textbf{Memory}, \textbf{Code}, and \textbf{Visual}. Fixations were parsed into transitions (e.g., Code $\rightarrow$ Memory) and aggregated into bigrams and trigrams to capture gaze sequences. To reduce noise, we applied smoothing filters to raw gaze coordinates, ignored fixations shorter than 100~ms~\cite{mahanama2022eye}, and windowed samples into 200~ms bins~\cite{niehorster2020characterizing}. Derived metrics (first fixation, transition frequencies, AOI dwell times) were used descriptively in our results to complement the qualitative analysis. Reproducible analysis code is available on OSF.\footnote{\url{https://osf.io/bdcj6/overview?view_only=e0170036e3a14200afc3823e2aa289c2}}

\subsection{Qualitative Analysis}

Our qualitative analysis followed reflexive thematic analysis~\cite{braun2021one} situated within an interpretivist epistemology. We treat participants' accounts not as objective reports about tool use but as situated narratives shaped by the interview setting, prior experiences, and disciplinary identities.

\paragraph{Analytic process.}
Our analysis was inductive with respect to RQ2: we did not enter with predetermined categories for what factors might shape engagement, but instead constructed themes through iterative coding. Two researchers independently engaged in repeated cycles of: (1)~\textit{Immersion}: reading and re-reading transcripts alongside screen recordings to attend to language, pauses, and interaction patterns; (2)~\textit{Open coding}: generating descriptive and interpretive codes capturing both what participants said (semantic level) and the assumptions or stances underlying those accounts (latent level); (3)~\textit{Memoing}: writing analytic memos to record emergent insights, tensions, and reflexive observations; (4)~\textit{Theme development}: clustering codes into higher-level patterns capturing both the affordances of each visualization and the affective and epistemic orientations participants adopted; and (5)~\textit{Iterative refinement}: revisiting transcripts to test coherence of emerging themes and attend to disconfirming cases.

\paragraph{Theoretical engagement.}
Frameworks such as MER~\cite{ainsworth2006deft} and CLT~\cite{paas2014cognitive, sweller2011cognitive} informed our later interpretation and positioning of findings, but did not guide initial coding. We entered the analysis aiming to allow patterns to emerge inductively, attentive to what existing theory might overlook.

\paragraph{Reflexivity.}
The lead analyst was a graduate student trained in qualitative methods, with research experience in visualization, computing education, and HCI, and extensive CS1/CS2 TA experience. Having contributed to the design of the MR visualizations, she engaged in repeated reflexive questioning to check for confirmation bias. A second researcher, an undergraduate with CS1/CS2 TA experience and proximity to the student experience, provided a complementary perspective. The team held collaborative reflection sessions with a broader research group including faculty with expertise in introductory computing instruction and visualization design. We acknowledge that our interpretations reflect our situated perspective as researchers invested in visualization design, and we approached analysis as a dialogic process.

Consistent with reflexive thematic analysis, we did not compute inter-rater reliability statistics. Braun and Clarke~\cite{braun2021one} argue that such measures presume a realist stance on coding that conflicts with RTA's interpretivist foundations. Instead, we treated multiple analysts as a resource for generating richer interpretations through dialogue rather than as a check on a single correct reading.
\section{Results}
\label{sec:results}

Despite carefully synchronized \Code, \Memory, and \Metaphor views designed to scaffold understanding, students spent nearly half their time (47\% median) focused on \Code, with some allocating over 70\% of attention to code even when visual representations were available. This section examines how students engaged with multi-view visualizations and what shaped their selective attention. We first present gaze analyses characterizing where students looked and how they transitioned between views, then draw on interview accounts to interpret the affective and social dimensions that shaped these patterns.

\subsection{Gaze Patterns Across Representations}
\label{sec:gaze}

\begin{figure*}[t]
  \centering
  \begin{subfigure}[t]{0.35\linewidth}
    \centering
    \includegraphics[width=\linewidth]{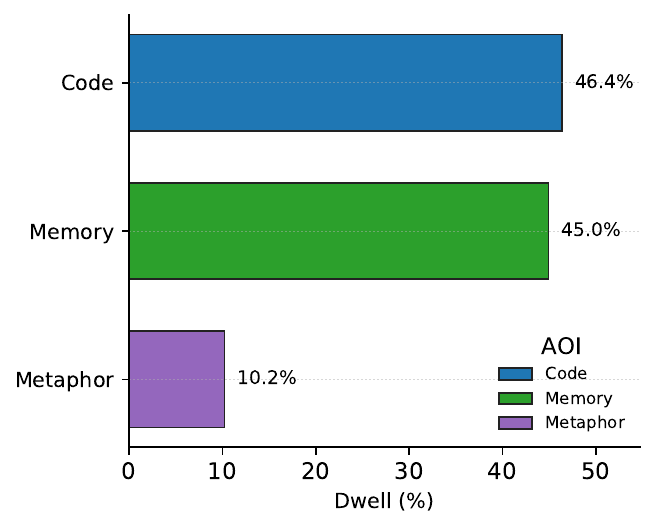}
    \caption{Overall proportion of dwell time in each area of interest
(AOI). Because values are medians across participants, they do not sum exactly to 100\%.}
   \Description{Bar chart showing the median proportion of gaze dwell time across three areas of interest in the multi-representational tool: Code (approximately 46\%), Memory (approximately 45\%), and Metaphor (approximately 10\%). Values represent medians across participants and do not sum exactly to 100 percent. Code and Memory receive substantially more attention than Metaphor.}
    \label{fig:gaze-distribution}
  \end{subfigure}
  \hfill
  \begin{subfigure}[t]{0.6\linewidth}
    \centering
    \includegraphics[width=\linewidth]{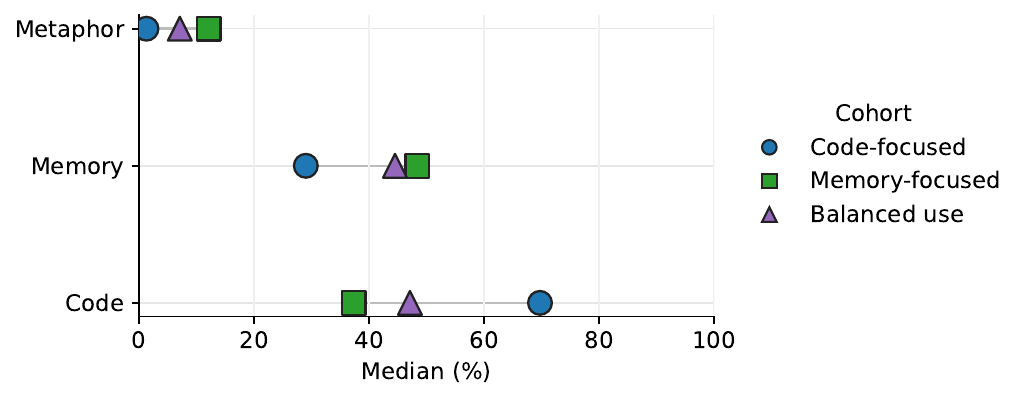}
    \caption{Median fixation distributions for three learner cohorts identified via k-means clustering on per-participant AOI dwell 
    percentages. Each marker shows a cohort's median dwell time on Code, Memory, and Metaphor panes; connecting lines indicate the 
    spread between cohorts on each AOI.}
    \Description{Scatter plot showing median fixation distributions for learner cohorts identified by clustering on area-of-interest dwell percentages. The horizontal axis shows median percentage of attention allocated to Code, Memory, and Metaphor. Cohorts include Code-focused, Memory-focused, and Balanced use groups. Points indicate that some learners concentrate heavily on Code, others emphasize Memory, and a smaller group distributes attention more evenly across panes.}
    \label{fig:cohort}
  \end{subfigure}
  \caption{Gaze fixations across panes in MER.}
  \label{fig:gaze-subfigures}
\end{figure*}

Fixation distributions varied systematically across representations and student experience (Figure~\ref{fig:gaze-subfigures}). Overall, \Code drew a median 47\% of attention, \Memory 34\%, and \Metaphor 9\%. This pattern was task-dependent: \Memory engagement peaked in linked-list tasks (47\% median), while \Metaphor gained traction in loop tasks (19\% median).

Students with prior experience (PE, $N{=}13$) showed more balanced distributions (\Code 39.6\%, \Memory 43.5\%, \Metaphor 13.2\%), whereas those without prior experience ($N{=}6$) anchored more heavily in \Code (48.1\%) with minimal \Metaphor use (3.3\%).  This is consistent with the suggestion from prior work that novices anchor in code traces while experienced learners coordinate multiple representations~\cite{aljehane2023studying}, and with evidence that higher intrinsic load during program comprehension manifests as more concentrated fixations on a single artifact~\cite{andrzejewska2020examining}. Effect sizes were moderate (Cliff's $\delta = -0.28$ to $0.33$) but underpowered given group sizes.

\begin{figure*}[t]
  \centering
  \includegraphics[width=0.85\linewidth]{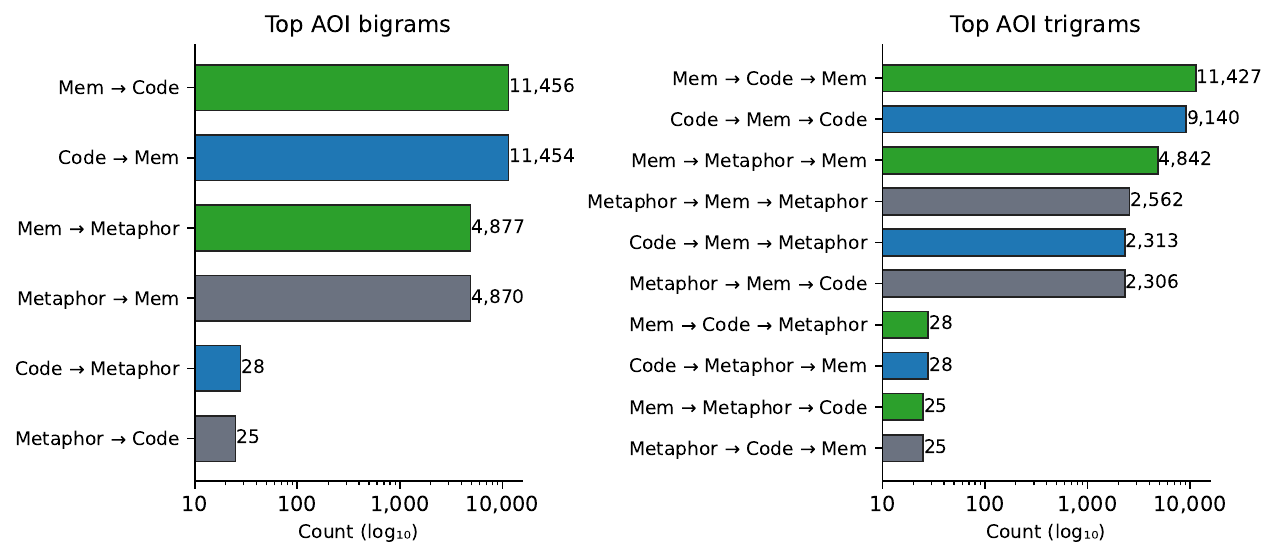}
  \caption{Transition frequencies across AOIs, showing integrative
cross-pane movement in MERs.}
  \Description{Two bar charts showing the most frequent gaze transitions between areas of interest in the multi-representational tool. The first chart displays top bigrams (two-step transitions), with the most frequent being Memory to Code and Code to Memory, each exceeding 11,000 occurrences. Transitions involving Metaphor occur less frequently. The second chart shows top trigrams (three-step transitions), with Memory–Code–Memory and Code–Memory–Code being most common. Counts are displayed on a logarithmic scale, indicating frequent integrative movement between Code and Memory panes and more selective use of Metaphor.}
  \label{fig:gaze-transitions}
\end{figure*}

Transition analyses revealed integrative cross-pane gaze movement rather than isolated viewing: \Memory$\leftrightarrow$\Code transitions exceeded 11k each, with \Metaphor selectively consulted in \Memory$\rightarrow$\Metaphor$\rightarrow$\Memory cycles (Figure~\ref{fig:gaze-transitions}). Most participants (12/19) began with \Memory, then cycled between \Code and \Memory while drawing in \Metaphor as needed. These results suggest that learners were not only viewing the panes but coordinating them, which supports abstraction~\cite{ainsworth2006deft} and representational flexibility~\cite{hill2015online,suh2007developing}.

Overall, gaze data show that learners treated \Code as primary while using MER’s tri-pane layout for integration. Students with prior experience used \Metaphor more often and switched between views more flexibly, while novices focused on code and returned to it repeatedly.

\subsection{Interview Accounts: What Shaped Selective Engagement}
\label{sec:interviews}

The gaze patterns above characterize \textit{how} students distributed attention (RQ1). We now turn to \textit{what shaped} that distribution (RQ2). Through open coding of interview transcripts, three themes were constructed that characterized variation across participants: students' relationship to cognitive effort and control (which we term \textit{agency}), individual differences in what felt helpful versus overwhelming (\textit{representational fit}), and concerns about what tool use signaled about academic identity (\textit{legitimacy}). We did not anticipate the social dimensions of the third theme; it was developed from repeated references to ``ego,'' ``childishness,'' and disciplinary appropriateness across multiple participants.

\subsubsection{Agency and Control Over Reasoning}
\label{sec:agency}

Students' \Code-dominant gaze patterns reflected not just habit but active agency-seeking. When asked why they focused on \Code despite available visual scaffolds, students framed manual tracing as essential to feeling in control of their own reasoning. $P6_A$ explained:

\begin{myquote2}[$P6_A$ (PE, BAB, Man)] 
I like tracing things on my own because I feel like I'm more in control. When I just press ``Next,'' I don't really know what's happening, I'm just kind of observing. But when you trace it yourself, you really have to understand how Python is going to execute the code.
\end{myquote2}

This preference for active engagement over guided observation appeared across experience levels. $P11_A$ went further, framing mental effort itself as valuable:

\begin{myquote2}[$P11_A$ (PE, ABA, Man)] 
With Python Tutor, you really have to get into the zone\ldots It takes effort and focus not to get overwhelmed\ldots Maybe there's something to be said for needing to think, because thinking might help you retain the knowledge better.
\end{myquote2}

$P11_A$ elaborated on this point at length, distinguishing between a mode where ``you're working harder to figure out what's going on'' and one where ``you can just sit back and let the visualizer guide your understanding more passively,'' arguing that the effortful mode ``helps you grow as a coder.''

Students also resisted function-level abstraction in the MR tool, where CS2 tasks stepped through execution at the function level rather than line by line. $P8_A$ argued that this compression undermined the tool's purpose:

\begin{myquote2}[$P8_A$ (PE, BAB, Man)] The problem with the memory visualization for linked lists is that it doesn't go line by line through what the append function actually does. It just executes the whole function and then shows the result. That kind of defeats the purpose of memory visualization.
\end{myquote2}

$P1_B$ echoed this, describing line-by-line pacing as ``especially useful for debugging.'' $P19_A$ connected this preference to her own practice: ``I'd only use [the memory visualization] if a question specifically asked me to draw it out. Otherwise, even if I were stuck on a problem, I wouldn't voluntarily start drawing out the full stack and everything.'' The issue was not accuracy (function-level views were correct) but whether abstraction meant skipping the work that participants associated with having a thorough understanding.

These accounts suggest that students valued visualizations that preserved their active role in reasoning. The high frequency of \Memory$\leftrightarrow$\Code transitions (11k+) supports this interpretation: students used visualizations for \textit{verification} of their own reasoning rather than as primary tools for understanding execution. As $P11_A$ put it, the visualization ``confirmed that my foundation was solid.'' $P10_A$ described a similar strategy: ``A lot was going on, so I started with the more abstract visual to get a basic sense, then used the memory pane to understand more in detail. Using them together worked really well.''

This pattern complicates a straightforward application of cognitive load theory to MER design. CLT emphasizes reducing extraneous load~\cite{sweller2011cognitive, paas2003cognitive}, but for these students, preserving active effort was more important than having that effort reduced. What matters is not just the cognitive demands of a representation but whether learners feel they control those demands.

\subsubsection{Individual Variation in Representational Fit}
\label{sec:calibration}

Low \Metaphor engagement (9\% median, 3.3\% for students without prior experience) accompanied wide variation in how students responded to the same visual designs. Some students described feeling overwhelmed by metaphorical elements. $P6_A$, who spent minimal time on \Metaphor, explained that he ``didn't really use it because it just looked too complicated.'' $P8_A$ elaborated on this visual overwhelm:

\begin{myquote2}[$P8_A$ (PE, BAB, Man)] 
I really don't like when there are just a bunch of arrows pointing everywhere. Once multiple things are happening at once, there are just too many arrows on the screen, and your brain has to keep track of where they all go\ldots There were just too many things going on, like a house, a chimney, people calling each other. It was a lot.
\end{myquote2}

$P19_A$ expressed a similar reaction but located the difficulty differently: ``the [metaphor] visual analogy could actually be more helpful [for beginners]\ldots But for more technical courses, I'd definitely say the memory analogy is the better option,'' suggesting that the issue was not the metaphor itself but its fit with her current level of understanding. $P5_B$ found the memory view more problematic than the metaphor: ``The memory analogy kind of looks confusing, because you don't really know what's calling what\ldots But in the [metaphor] visual, it's like: this is your house, and this is what you have access to.''

Yet other students described the same metaphorical representations as essential. $P4_B$ connected this directly to how he learns:

\begin{myquote2}[$P4_B$ (PE, BAB, Man)] 
In computer science, some concepts are really abstract, and to understand them you need to connect them to something you've already experienced. Lockers for memory models, supermarket shelves for searching. Honestly, if it wasn't for that, I don't think I would have really grasped the concept.
\end{myquote2}

$P13_A$ described a similar reliance on visual scaffolds: ``This visualization is much more helpful. I remember using it on the course website, and I just felt more comfortable with it\ldots Even though the visual might feel a bit cartoonish, I still think it's more effective because of how clearly it explains each step.'' $P15_A$ found the visualizations valuable specifically for complex topics: ``it helps me with understanding the more complex ones, where it's like, sometimes you understand the general concept, but not the specific thing.''

Students also validated representations through resemblance to their own existing practices. $P14_A$ noted that the MR tool's memory view ``does all the crossing out'' she would do manually when tracing, and $P12_A$ described it as matching what he would ``draw out'' himself. $P9_A$ put it simply: ``this is exactly what I would draw out [on a test].'' When visualizations matched familiar practices, they felt natural; when they departed from those practices, even in pedagogically motivated ways, students found them harder to adopt.

This variation reveals that what counts as ``concrete'' or ``overwhelming'' is not a property of the representation alone but depends on the learner's interpretive capacity, prior habits, and current goals. A visualization that can scaffold understanding for one student can also create additional interpretive burden for another.

\subsubsection{Legitimacy and Disciplinary Positioning}
\label{sec:legitimacy}

Low \Metaphor engagement also reflected social positioning. Students explicitly rejected metaphors as inappropriate for university-level work, even when acknowledging potential pedagogical value. $P18_A$ articulated this tension directly:

\begin{myquote2}[$P18_A$ (noPE, BAB, Man)] As a university student, using these kind of graphics is kind of\ldots definitely not demeaning, but kind of almost a little bit childish. But then again, it might just be like ego and personal bias.
\end{myquote2}

$P1_B$ echoed this framing: ``when there's a bunny and a magician it kind of feels like something that would be elementary\ldots That's kind of like an ego thing.'' Strikingly, both students explicitly named ``ego,'' recognizing that their resistance was not about the visualization's pedagogical quality but about how using it positioned them. $P17_A$ similarly described the metaphor view as ``definitely good for younger people'' and said his younger brother ``might enjoy this type of interface,'' distancing himself from the intended audience. $P2_B$ framed MER as ``for CS1'' and ``for someone who's never coded before,'' implicitly claiming advanced status.

This positioning had consequences for engagement. $P18_A$ was a student without prior experience who could have benefited from visual scaffolds, yet avoided \Metaphor (10.4\% dwell time) to signal seriousness. The concern was not comprehension but identity: what does it mean to use ``childish'' tools in a university computer science course?

Legitimacy concerns also shaped preferences for technical detail. Students equated exhaustive tracing with real understanding and technical fidelity with rigor. $P8_A$ insisted that memory visualization should show ``every single step'' because ``the whole purpose of memory visualization is to get really nitty-gritty, step by step, about what happens in memory.'' $P13_A$ described the memory view as matching what ``we did in [CS2],'' connecting tool legitimacy to course practices. $P2_B$ explained his preference for Python Tutor by noting that ``it talks about different frames and how objects are being created\ldots it's more structured,'' valuing technical vocabulary and formal structure over visual accessibility.

However, this pattern was not universal. $P4_B$ (23.9\% \Metaphor dwell time, nearly double the median) embraced analogies as central to his learning: ``I always think in analogies.'' $P1_B$ found the visual analogy for scope ``actually really well thought out'' and ``more helpful'' despite his earlier legitimacy concerns about the loop metaphor. And $P7_A$ described the MR tool's memory view as preferable specifically because it was ``more helpful for children'' and ``more engaging,'' suggesting that what some students rejected as childish, others valued as accessible.

These accounts reveal that students' engagement with visualizations was shaped not only by cognitive fit but by what tool use signaled about their identity as computer science students. Representations that felt ``too easy'' or ``too playful'' were avoided even when they might have helped, because the social cost of appearing unserious outweighed the potential learning benefit. This is distinct from the individual variation described in the previous section: a student might find a metaphor personally helpful yet still avoid it because it feels illegitimate in a university context.

\section{Discussion}
\label{sec:discussion}

Our study set out to understand what factors shape whether students engage with or resist multi-representational visualizations when learning programming. Through gaze tracking and interviews with 19 students using synchronized {\Code}, {\Memory}, and {\Metaphor} views, we identified three themes that characterized variation in engagement: students' need to control their own cognitive effort (agency), individual differences in what counts as helpful versus overwhelming (representational fit), and social concerns about what tools signal regarding academic identity (legitimacy). In this section, we situate these themes within broader literature on learning, discuss what they mean for how we deploy visualization tools in computing education, and note limitations of our work.

\subsection{Agency and the Role of Effort in Learning}

Students consistently anchored reasoning in {\Code}, framing manual tracing as the ``real'' work of programming and describing automated stepping as passive observation. They were not simply unaware of the visual scaffolds; they actively avoided them to preserve control over cognitive effort.

This finding resonates with research on desirable difficulties~\cite{bjork2011making} and productive struggle~\cite{young2024productive}, which shows that effortful generation can create more durable learning than smooth consumption. However, our participants drew a distinction that this literature does not always make explicit: they wanted \textit{control over} difficulty, not difficulty per se. When tools automated stepping, students experienced a loss of agency rather than a reduction in load. When they chose to trace manually and then checked their reasoning against a visualization, the same cognitive work felt productive. This aligns with findings from \citet{kirk2019perceiving} and \citet{klepsch2021making}, who show that learners’ subjective interpretations of effort shape both their strategy choices and their load appraisals. Effort experienced as actively invested aligns with germane load, whereas effort experienced as imposed or strenuous aligns more closely with intrinsic load and is often interpreted as reduced learning.

Research on strategic friction~\cite{cox2016design}, uncomfortable interactions~\cite{benford2012uncomfortable}, and ambiguity in design~\cite{gaver2003ambiguity} has shown that small obstacles can prompt reflection and deeper engagement. Our findings suggest a parallel principle for educational visualization: friction that learners control can feel productive, while friction that tools impose can feel alienating. The students who engaged most successfully with our MR tool were those who used visualizations as \textit{verification instruments} rather than explanation guides, maintaining their own reasoning while selectively consulting views for confirmation.

For computing education, this has practical consequences. When we deploy visualization tools in courses, the default instinct is to reduce friction: automate stepping, simplify representations, guide attention. Our findings suggest that for some students, this removes the engagement that makes the tool feel worthwhile. Tools that support where students choose when to reveal visual supports rather than having them presented automatically, may better preserve the sense of active construction that students associate with learning. This does not mean abandoning scaffolding, but rather designing scaffolds that students can opt into rather than having imposed on them.

\subsection{No Universal ``Concrete''}

The same metaphorical representations were essential for some students yet overwhelming for others, despite identical designs and similar experience levels. $P4_B$ described metaphors as foundational to understanding, while $P8_A$ found even {\Memory} arrows overwhelming. $P5_B$ found the memory view more confusing than the metaphor, while $P2_B$ had exactly the opposite reaction.

This variation complicates the assumptions of concreteness fading theory~\cite{fyfe2014concreteness}, which posits a universal progression from concrete to abstract representations. In our sample, what counted as ``concrete'' (accessible, grounding) for one student was ``cluttered'' (overwhelming, distracting) for another. This aligns with Wong et al.'s finding that tool preferences manifest along continua rather than discrete stages~\cite{wong2025spectrum}, and suggests the expertise reversal effect may operate not only on the amount of instructional support but on which representational form a given learner experiences as grounding~\cite{kalyuga2007expertise}. More experienced students (those with prior programming experience, PE; see Section~\ref{subsec:participants}) stated preferences for technical fidelity over metaphorical approachability, yet their gaze data revealed they actually engaged {\Metaphor} substantially in linked-list tasks (13.2\% median vs.\ 3.3\% for students without prior experience). This gap between stated preference and observed behavior suggests that experienced learners may selectively deploy representations they publicly frame as beneath them.

Students also validated representations through resemblance to their existing practices~\cite{reber2017processing, scheiter2020effort}. When {\Memory} matched diagrams they would draw on paper, it felt natural and effective. This creates a tension for tool design as mirroring learners' existing habits reduces friction and increases acceptance, but it can also reinforce surface-level strategies. Representations that differ from familiar forms create disfluency that, if appropriately scaffolded, can prompt productive reflection~\cite{gaver2003ambiguity}. The challenge for instructors and tool designers is determining when familiarity enables progress and when it constrains learners within existing mental models, a calibration that likely shifts as students develop expertise.

For course design, this suggests that one-size-fits-all deployment of visualization tools may be less effective than providing options. Toggleable abstraction levels, where students can adjust visual density to their current needs, would allow the same tool to serve students who find metaphors essential and those who find them overwhelming.

\subsection{Legitimacy and Disciplinary Identity}

Many students explicitly rejected \Metaphor as inappropriate for university-level computer science, naming ``ego'' directly and distancing themselves from representations they deemed ``childish'' or suitable for ``beginners.'' This resistance occurred even among students without prior experience who stood to benefit most from visual scaffolds: $P18_A$ avoided {\Metaphor} to signal seriousness despite being relatively new to programming.

This pattern connects to broader work on identity and belonging in computing education. Gee's~\cite{gee2000chapter} framework of identity as social positioning suggests that students are not just learning content but learning to \textit{be} computer scientists, and tool choices become part of that identity work. Nasir and Hand~\cite{nasir2006exploring} show how learning settings shape which identities feel available and valued. In our study, students who equated technical detail with rigor and playful metaphors with novice status were performing a particular version of what it means to be a serious computer science student. This is related to but distinct from the representational fit discussed above: a student might find a metaphor personally helpful yet still avoid it because using it feels illegitimate in a university context.

This finding also resonates with Legitimation Code Theory~\cite{maton2015knowledge}, which examines how forms of knowledge are valued differently across contexts. In university computer science, technical detail signals competence, while playful scaffolds may signal immaturity, regardless of their actual learning benefits. For instructors, this means that \textit{how} a tool is framed may matter as much as \textit{how} it is designed. Presenting the same visualization as a ``professional debugging tool'' versus a ``learning aid'' could shift whether students feel it is appropriate for them. Similarly, when instructors model the use of visual representations in their own problem-solving, they signal that such tools are part of expert practice.

\subsection{Interactions Across Themes}

These three themes operated interactively in our data. Students who engaged most successfully with the MR tool were those for whom all three themes aligned: they maintained agency by anchoring reasoning in code while selectively consulting views, found individual fit with the available representations, and framed technical detail as appropriately rigorous. When any single theme was violated, even well-designed views were rejected. Students avoided {\Metaphor} when it threatened legitimacy, resisted function-level abstraction when it removed control, and dismissed metaphors that overwhelmed them even when the same designs helped peers.

This helps explain why MER tools that embody sound cognitive design principles sometimes fail to gain traction with students~\cite{schwonke2009multiple}. Prior CER work established that active engagement, not visual representation design, predicts learning outcomes~\cite{hundhausen2002meta, naps2002exploring}. Our findings extend that insight by showing that engagement itself is not reducible to interaction design. Even when tools offer the forms of activity that engagement taxonomies recommend (responding, changing, exploring), students may decline to take them up for reasons that cognitive design principles may not anticipate. MER theory's cognitive principles of complementarity, constraint, and construction~\cite{ainsworth2006deft} are important for representational design, but they do not predict engagement when affective and social factors are misaligned. A visualization can be complementary (showing both code and memory state), constraining (guiding attention to key relationships), and supportive of construction (enabling translation between views), yet still be avoided if it violates a student's need for agency, incorrectly estimates their interpretive capacity, or threatens their disciplinary positioning.

\subsection{Limitations and Scope}
\label{sec:limitations}

Our visualization tool was a research probe, the specific metaphors we chose inevitably shaped student reactions, and different visual designs might elicit different responses. Our 19 participants participated from two campuses of one university, all having recently completed CS1 and CS2 in Python. Although we included variation in prior experience, gender, and campus, participants share a common institutional context. We offer our themes as analytically grounded characterizations, not universal claims. Establishing transferability requires future work. The themes we identified are not necessarily exhaustive. Factors such as assessment pressure, peer dynamics, or sustained use over a full term may matter in ways a single-session tool study cannot capture. We did not analyze task accuracy or completion rates as outcomes. While this was consistent with our focus on engagement rather than learning outcomes, it means we cannot directly link the engagement patterns we observed to performance differences.

Our webcam-based gaze tracking cannot support fine-grained attention claims~\cite{papoutsaki2018eye, slim2023moving}. We treat gaze data descriptively, as a complement to interview accounts. 

\subsection{Implications for Teaching and Tool Design}

Our findings suggest several practical considerations. First, visualization tools may be more effective when positioned as on-demand checking mechanisms rather than step-by-step guides. Prediction-based workflows, where students commit to an expected state before revealing the visualization, have a long history in CER~\cite{naps2002exploring, hundhausen2002meta} and align well with what our participants described as productive use. Our contribution is not to reintroduce prediction as a mechanism but to suggest that its value may lie partly in how it preserves learner agency, a framing complementary to earlier engagement-centered accounts and to uses of visualization in algorithm courses~\cite{velazquez2016srec}. Second, providing toggleable abstraction levels would allow students to calibrate visual density to their current capacity, as settings that work for one student can overwhelm another. Prior work has long advocated multiple abstraction levels~\cite{hansen2002designing} and InfoVis-style filter and detail-on-demand techniques for program animation~\cite{velazquez2010infovis}. Our findings add a learner-side rationale as the same toggling mechanisms can serve as agency supports, letting students opt into density rather than having it imposed. Third, how tools are introduced matters: instructors who model visualization use in their own reasoning signal that these tools are part of expert practice, and providing appropriate framing (e.g., ``technical notation'' vs.\ ``conceptual scaffolds'') allows students to engage without threatening their disciplinary identity. Finally, synchronized highlighting across coordinated views has been a staple of program animation~\cite{velazquez2010infovis, hansen2002designing}. What our data suggest is that these linking mechanisms serve students best when used as verification instruments, tools for checking self-generated reasoning, rather than as guided walkthroughs that students follow passively. This is a shift in stance rather than in mechanism, and it aligns with the engagement taxonomy's emphasis that what matters is not the feature but the activity it supports~\cite{naps2002exploring}.
\section{Conclusion}

Our study examined how 19 undergraduates engaged with multi-representational visualizations linking {\Code}, {\Memory}, and {\Metaphor} views across core CS1 and CS2 topics. Through gaze tracking and reflective interviews, we identified 3 themes that shaped whether and how students engaged with these representations: agency (students' need to control their own cognitive effort rather than have it reduced), representational fit (wide individual variation in what counts as helpful versus overwhelming), and legitimacy (social concerns about what tools signal regarding academic identity).

These themes help explain why well-designed program visualizations sometimes fail to gain traction with students. In our study, identical designs were essential scaffolds for some students yet sources of frustration or social discomfort for others. Automation intended to reduce cognitive load was resisted because it felt passive, and metaphors designed to make abstraction accessible were avoided because they felt childish. Students did not simply use or ignore visualizations; they adapted, resisted, and reinterpreted them in ways that reflected their prior experiences, learning strategies, and sense of who they were becoming as computer scientists.

For computing education research, our findings suggest that evaluating visualization tools on cognitive design alone may miss consequential sources of variation in student engagement. Attending to how tools preserve learner autonomy, how they calibrate to individual differences, and how they are framed within disciplinary contexts may be as important as attending to representational complementarity or cognitive load. For instructors, practical steps include positioning visualizations as verification tools rather than explanation guides, providing representational choice so students can calibrate complexity to their needs, and modeling visualization use to signal its legitimacy as part of expert practice.

Our themes are grounded in one study at one institution, and their scope and relative importance in other settings remain open questions. We hope they provide a useful starting point for future work investigating the affective and social dimensions of tool engagement in programming education.

\begin{acks}
We thank Valeria Ramirez Osorio for hand-drawing and designing the visual materials used in the CS1 course content. We also acknowledge the support of the Natural Sciences and Engineering Research Council of Canada (NSERC) Discovery Grant \#RGPIN-2024-04348, RGPIN-2024-04348, PGS D–600673–2025, and USRA-682989.
Cette recherche a été financée par le Conseil de recherches en sciences naturelles et en génie du Canada (CRSNG), subvention à la découverte RGPIN-2024-04348, RGPIN-2024-04348, PGS D-600673-2025, et USRA-682989.
\end{acks}

\bibliographystyle{ACM-Reference-Format}
\bibliography{references}


\begin{thebibliography}{112}


\ifx \showCODEN    \undefined \def \showCODEN     #1{\unskip}     \fi
\ifx \showISBNx    \undefined \def \showISBNx     #1{\unskip}     \fi
\ifx \showISBNxiii \undefined \def \showISBNxiii  #1{\unskip}     \fi
\ifx \showISSN     \undefined \def \showISSN      #1{\unskip}     \fi
\ifx \showLCCN     \undefined \def \showLCCN      #1{\unskip}     \fi
\ifx \shownote     \undefined \def \shownote      #1{#1}          \fi
\ifx \showarticletitle \undefined \def \showarticletitle #1{#1}   \fi
\ifx \showURL      \undefined \def \showURL       {\relax}        \fi
\providecommand\bibfield[2]{#2}
\providecommand\bibinfo[2]{#2}
\providecommand\natexlab[1]{#1}
\providecommand\showeprint[2][]{arXiv:#2}

\bibitem[Adolph and Kretch(2015)]%
        {adolph2015gibson}
\bibfield{author}{\bibinfo{person}{Karen~E Adolph} {and} \bibinfo{person}{Kari~S Kretch}.} \bibinfo{year}{2015}\natexlab{}.
\newblock \showarticletitle{Gibson’s Theory of Perceptual Learning}.
\newblock \bibinfo{journal}{\emph{International Encyclopedia of the Social and Behavioral Sciences}}  \bibinfo{volume}{10} (\bibinfo{year}{2015}), \bibinfo{pages}{127--134}.
\newblock
\href{https://doi.org/10.1016/b978-0-08-097086-8.23096-1}{doi:\nolinkurl{10.1016/b978-0-08-097086-8.23096-1}}


\bibitem[Aedo~Lopez et~al\mbox{.}(2016)]%
        {aedo2016teaching}
\bibfield{author}{\bibinfo{person}{Marco Aedo~Lopez}, \bibinfo{person}{Elizabeth Vidal~Duarte}, \bibinfo{person}{Eveling Castro~Gutierrez}, {and} \bibinfo{person}{Alfredo Paz~Valderrama}.} \bibinfo{year}{2016}\natexlab{}.
\newblock \showarticletitle{Teaching Abstraction, Function and Reuse in the First Class of CS1: A Lightbot Experience}. In \bibinfo{booktitle}{\emph{Proceedings of the 2016 ACM Conference on Innovation and Technology in Computer Science Education}}. \bibinfo{publisher}{ACM}, \bibinfo{address}{New York, NY, USA}, \bibinfo{pages}{256--257}.
\newblock
\href{https://doi.org/10.1145/2899415.2925505}{doi:\nolinkurl{10.1145/2899415.2925505}}


\bibitem[Ahn et~al\mbox{.}(2020)]%
        {ahn2020towards}
\bibfield{author}{\bibinfo{person}{Seoyoung Ahn}, \bibinfo{person}{Conor Kelton}, \bibinfo{person}{Aruna Balasubramanian}, {and} \bibinfo{person}{Greg Zelinsky}.} \bibinfo{year}{2020}\natexlab{}.
\newblock \showarticletitle{Towards Predicting Reading Comprehension From Gaze Behavior}. In \bibinfo{booktitle}{\emph{ACM Symposium on Eye Tracking Research and Applications}}. \bibinfo{publisher}{ACM}, \bibinfo{address}{New York, NY, USA}, \bibinfo{pages}{1--5}.
\newblock
\href{https://doi.org/10.1145/3379156.3391335}{doi:\nolinkurl{10.1145/3379156.3391335}}


\bibitem[Ainsworth(1999)]%
        {ainsworth_functions_1999}
\bibfield{author}{\bibinfo{person}{Shaaron Ainsworth}.} \bibinfo{year}{1999}\natexlab{}.
\newblock \showarticletitle{The Functions of Multiple Representations}.
\newblock \bibinfo{journal}{\emph{Computers \& Education}} \bibinfo{volume}{33}, \bibinfo{number}{2} (\bibinfo{date}{September} \bibinfo{year}{1999}), \bibinfo{pages}{131--152}.
\newblock
\showISSN{0360-1315}
\href{https://doi.org/10.1016/S0360-1315(99)00029-9}{doi:\nolinkurl{10.1016/S0360-1315(99)00029-9}}


\bibitem[Ainsworth(2006)]%
        {ainsworth2006deft}
\bibfield{author}{\bibinfo{person}{Shaaron Ainsworth}.} \bibinfo{year}{2006}\natexlab{}.
\newblock \showarticletitle{DeFT: A Conceptual Framework for Considering Learning With Multiple Representations}.
\newblock \bibinfo{journal}{\emph{Learning and Instruction}} \bibinfo{volume}{16}, \bibinfo{number}{3} (\bibinfo{year}{2006}), \bibinfo{pages}{183--198}.
\newblock
\href{https://doi.org/10.1016/j.learninstruc.2006.03.001}{doi:\nolinkurl{10.1016/j.learninstruc.2006.03.001}}


\bibitem[Aljehane et~al\mbox{.}(2023)]%
        {aljehane2023studying}
\bibfield{author}{\bibinfo{person}{Salwa~D Aljehane}, \bibinfo{person}{Bonita Sharif}, {and} \bibinfo{person}{Jonathan~I Maletic}.} \bibinfo{year}{2023}\natexlab{}.
\newblock \showarticletitle{Studying Developer Eye Movements to Measure Cognitive Workload and Visual Effort for Expertise Assessment}.
\newblock \bibinfo{journal}{\emph{Proceedings of the ACM on Human-Computer Interaction}} \bibinfo{volume}{7}, \bibinfo{number}{ETRA} (\bibinfo{year}{2023}), \bibinfo{pages}{1--18}.
\newblock
\href{https://doi.org/10.1145/3591135}{doi:\nolinkurl{10.1145/3591135}}


\bibitem[Almadhoun and Parham-Mocello(2021)]%
        {almadhoun2021exploratory}
\bibfield{author}{\bibinfo{person}{Eman Almadhoun} {and} \bibinfo{person}{Jennifer Parham-Mocello}.} \bibinfo{year}{2021}\natexlab{}.
\newblock \showarticletitle{Exploratory Study on Accuracy of Students' Mental Models of a Singly Linked List}. In \bibinfo{booktitle}{\emph{2021 IEEE Frontiers in Education Conference (FIE)}}. \bibinfo{publisher}{IEEE}, \bibinfo{address}{Piscataway, NJ, USA}, \bibinfo{pages}{1--9}.
\newblock
\href{https://doi.org/10.1109/fie49875.2021.9637318}{doi:\nolinkurl{10.1109/fie49875.2021.9637318}}


\bibitem[Andrzejewska and Skawi{\'n}ska(2020)]%
        {andrzejewska2020examining}
\bibfield{author}{\bibinfo{person}{Magdalena Andrzejewska} {and} \bibinfo{person}{Agnieszka Skawi{\'n}ska}.} \bibinfo{year}{2020}\natexlab{}.
\newblock \showarticletitle{Examining Students’ Intrinsic Cognitive Load During Program Comprehension--an Eye Tracking Approach}. In \bibinfo{booktitle}{\emph{International Conference on Artificial Intelligence in Education}}. Springer, \bibinfo{publisher}{Springer International Publishing}, \bibinfo{address}{Berlin, Germany}, \bibinfo{pages}{25--30}.
\newblock
\href{https://doi.org/10.1007/978-3-030-52240-7_5}{doi:\nolinkurl{10.1007/978-3-030-52240-7_5}}


\bibitem[Baldonado et~al\mbox{.}(2000)]%
        {baldonado2000guidelines}
\bibfield{author}{\bibinfo{person}{Michelle Q~Wang Baldonado}, \bibinfo{person}{Allison Woodruff}, \bibinfo{person}{Allan Kuchinsky}, {et~al\mbox{.}}} \bibinfo{year}{2000}\natexlab{}.
\newblock \showarticletitle{Guidelines for Using Multiple Views in Information Visualization.}. In \bibinfo{booktitle}{\emph{Advanced Visual Interfaces}}, Vol.~\bibinfo{volume}{10}. \bibinfo{publisher}{ACM}, \bibinfo{address}{New York, NY, USA}, \bibinfo{pages}{345513--345271}.
\newblock
\href{https://doi.org/10.1145/345513.345271}{doi:\nolinkurl{10.1145/345513.345271}}


\bibitem[Balijepally et~al\mbox{.}(2012)]%
        {balijepally2012effect}
\bibfield{author}{\bibinfo{person}{VenuGopal Balijepally}, \bibinfo{person}{Sridhar Nerur}, {and} \bibinfo{person}{RadhaKanta Mahapatra}.} \bibinfo{year}{2012}\natexlab{}.
\newblock \showarticletitle{Effect of Task Mental Models on Software Developer's Performance: An Experimental Investigation}. In \bibinfo{booktitle}{\emph{2012 45th Hawaii International Conference on System Sciences}}. \bibinfo{publisher}{IEEE}, \bibinfo{address}{Piscataway, NJ, USA}, \bibinfo{pages}{5442--5451}.
\newblock
\href{https://doi.org/10.1109/hicss.2012.9}{doi:\nolinkurl{10.1109/hicss.2012.9}}


\bibitem[Bansal et~al\mbox{.}(2021)]%
        {bansal2021eye}
\bibfield{author}{\bibinfo{person}{Aman Bansal}, \bibinfo{person}{Preey Shah}, {and} \bibinfo{person}{Sahil Shah}.} \bibinfo{year}{2021}\natexlab{}.
\newblock \bibinfo{title}{Eye: Program Visualizer for CS2}.
\newblock
\showeprint[arxiv]{2101.12089}~[cs.CY]
\urldef\tempurl%
\url{https://arxiv.org/abs/2101.12089}
\showURL{%
\tempurl}


\bibitem[Bednarik(2012)]%
        {bednarik2012expertise}
\bibfield{author}{\bibinfo{person}{Roman Bednarik}.} \bibinfo{year}{2012}\natexlab{}.
\newblock \showarticletitle{Expertise-Dependent Visual Attention Strategies Develop Over Time During Debugging With Multiple Code Representations}.
\newblock \bibinfo{journal}{\emph{International Journal of Human-Computer Studies}} \bibinfo{volume}{70}, \bibinfo{number}{2} (\bibinfo{year}{2012}), \bibinfo{pages}{143--155}.
\newblock
\href{https://doi.org/10.1016/j.ijhcs.2011.09.003}{doi:\nolinkurl{10.1016/j.ijhcs.2011.09.003}}


\bibitem[Benford et~al\mbox{.}(2012)]%
        {benford2012uncomfortable}
\bibfield{author}{\bibinfo{person}{Steve Benford}, \bibinfo{person}{Chris Greenhalgh}, \bibinfo{person}{Gabriella Giannachi}, \bibinfo{person}{Brendan Walker}, \bibinfo{person}{Joe Marshall}, {and} \bibinfo{person}{Tom Rodden}.} \bibinfo{year}{2012}\natexlab{}.
\newblock \showarticletitle{Uncomfortable Interactions}. In \bibinfo{booktitle}{\emph{Proceedings of the SIGCHI Conference on Human Factors in Computing Systems}}. \bibinfo{publisher}{ACM}, \bibinfo{address}{New York, NY, USA}, \bibinfo{pages}{2005--2014}.
\newblock
\href{https://doi.org/10.4324/9780429242816-8}{doi:\nolinkurl{10.4324/9780429242816-8}}


\bibitem[Bjork et~al\mbox{.}(2011)]%
        {bjork2011making}
\bibfield{author}{\bibinfo{person}{Elizabeth~L Bjork}, \bibinfo{person}{Robert~A Bjork}, {et~al\mbox{.}}} \bibinfo{year}{2011}\natexlab{}.
\newblock \showarticletitle{Making Things Hard on Yourself, but in a Good Way: Creating Desirable Difficulties to Enhance Learning}.
\newblock \bibinfo{journal}{\emph{Psychology and the Real World: Essays Illustrating Fundamental Contributions to Society}} \bibinfo{volume}{2}, \bibinfo{number}{59-68} (\bibinfo{year}{2011}), \bibinfo{pages}{56--64}.
\newblock


\bibitem[Braun and Clarke(2021)]%
        {braun2021one}
\bibfield{author}{\bibinfo{person}{Virginia Braun} {and} \bibinfo{person}{Victoria Clarke}.} \bibinfo{year}{2021}\natexlab{}.
\newblock \showarticletitle{One Size Fits All? What Counts as Quality Practice in (Reflexive) Thematic Analysis?}
\newblock \bibinfo{journal}{\emph{Qualitative Research in Psychology}} \bibinfo{volume}{18}, \bibinfo{number}{3} (\bibinfo{year}{2021}), \bibinfo{pages}{328--352}.
\newblock
\href{https://doi.org/10.1080/14780887.2020.1769238}{doi:\nolinkurl{10.1080/14780887.2020.1769238}}


\bibitem[Brooks(1978)]%
        {brooks1978using}
\bibfield{author}{\bibinfo{person}{Ruven Brooks}.} \bibinfo{year}{1978}\natexlab{}.
\newblock \showarticletitle{Using a Behavioral Theory of Program Comprehension in Software Engineering}. In \bibinfo{booktitle}{\emph{Proceedings of the 3rd International Conference on Software Engineering}}. \bibinfo{publisher}{ACM}, \bibinfo{address}{New York, NY, USA}, \bibinfo{pages}{196--201}.
\newblock


\bibitem[Brown and Sedgewick(1984)]%
        {brown1984system}
\bibfield{author}{\bibinfo{person}{Marc~H Brown} {and} \bibinfo{person}{Robert Sedgewick}.} \bibinfo{year}{1984}\natexlab{}.
\newblock \showarticletitle{A System for Algorithm Animation}. In \bibinfo{booktitle}{\emph{Proceedings of the 11th Annual Conference on Computer Graphics and Interactive Techniques}}. \bibinfo{publisher}{ACM}, \bibinfo{address}{New York, NY, USA}, \bibinfo{pages}{177--186}.
\newblock
\href{https://doi.org/10.1145/964965.808596}{doi:\nolinkurl{10.1145/964965.808596}}


\bibitem[Brusilovsky(1993)]%
        {brusilovsky1993program}
\bibfield{author}{\bibinfo{person}{Peter Brusilovsky}.} \bibinfo{year}{1993}\natexlab{}.
\newblock \showarticletitle{Program Visualization as a Debugging Tool for Novices}. In \bibinfo{booktitle}{\emph{INTERACT'93 and CHI'93 Conference Companion on Human Factors in Computing Systems}}. \bibinfo{publisher}{ACM Press}, \bibinfo{address}{New York, NY, USA}, \bibinfo{pages}{29--30}.
\newblock
\href{https://doi.org/10.1145/259964.260031}{doi:\nolinkurl{10.1145/259964.260031}}


\bibitem[Ca{\~n}as et~al\mbox{.}(2001)]%
        {canas2001role}
\bibfield{author}{\bibinfo{person}{Jos{\'e}~J Ca{\~n}as}, \bibinfo{person}{Adoraci{\'o}n Antol{\'\i}}, {and} \bibinfo{person}{Jos{\'e}~F Quesada}.} \bibinfo{year}{2001}\natexlab{}.
\newblock \showarticletitle{The Role of Working Memory on Measuring Mental Models of Physical Systems}.
\newblock \bibinfo{journal}{\emph{Psicol{\'o}gica}} \bibinfo{volume}{22}, \bibinfo{number}{1} (\bibinfo{year}{2001}), \bibinfo{pages}{25--42}.
\newblock


\bibitem[Card et~al\mbox{.}(1999)]%
        {card1999readings}
\bibfield{author}{\bibinfo{person}{Stuart~K Card}, \bibinfo{person}{Jock Mackinlay}, {and} \bibinfo{person}{Ben Shneiderman}.} \bibinfo{year}{1999}\natexlab{}.
\newblock \bibinfo{booktitle}{\emph{Readings in Information Visualization: Using Vision to Think}}.
\newblock \bibinfo{publisher}{Morgan Kaufmann}, \bibinfo{address}{San Francisco, CA, USA}.
\newblock


\bibitem[Cherenkova et~al\mbox{.}(2014)]%
        {cherenkova2014identifying}
\bibfield{author}{\bibinfo{person}{Yuliya Cherenkova}, \bibinfo{person}{Daniel Zingaro}, {and} \bibinfo{person}{Andrew Petersen}.} \bibinfo{year}{2014}\natexlab{}.
\newblock \showarticletitle{Identifying Challenging CS1 Concepts in a Large Problem Dataset}. In \bibinfo{booktitle}{\emph{Proceedings of the 45th ACM Technical Symposium on Computer Science Education}}. \bibinfo{publisher}{ACM}, \bibinfo{address}{New York, NY, USA}, \bibinfo{pages}{695--700}.
\newblock
\href{https://doi.org/10.1145/2538862.2538966}{doi:\nolinkurl{10.1145/2538862.2538966}}


\bibitem[Clancy(2005)]%
        {clancy2005misconceptions}
\bibfield{author}{\bibinfo{person}{Michael Clancy}.} \bibinfo{year}{2005}\natexlab{}.
\newblock \showarticletitle{Misconceptions and Attitudes That Interfere With Learning to Program}.
\newblock In \bibinfo{booktitle}{\emph{Computer Science Education Research}}. \bibinfo{publisher}{Taylor \& Francis}, \bibinfo{address}{London, UK}, \bibinfo{pages}{95--110}.
\newblock
\href{https://doi.org/10.1201/9781482287325-18}{doi:\nolinkurl{10.1201/9781482287325-18}}


\bibitem[Cox et~al\mbox{.}(2016)]%
        {cox2016design}
\bibfield{author}{\bibinfo{person}{Anna~L Cox}, \bibinfo{person}{Sandy~JJ Gould}, \bibinfo{person}{Marta~E Cecchinato}, \bibinfo{person}{Ioanna Iacovides}, {and} \bibinfo{person}{Ian Renfree}.} \bibinfo{year}{2016}\natexlab{}.
\newblock \showarticletitle{Design Frictions for Mindful Interactions: The Case for Microboundaries}. In \bibinfo{booktitle}{\emph{Proceedings of the 2016 CHI Conference Extended Abstracts on Human Factors in Computing Systems}}. \bibinfo{publisher}{ACM}, \bibinfo{address}{New York, NY, USA}, \bibinfo{pages}{1389--1397}.
\newblock


\bibitem[Daniel et~al\mbox{.}(2018)]%
        {daniel2018representational}
\bibfield{author}{\bibinfo{person}{Kristy~L Daniel}, \bibinfo{person}{Carrie~Jo Bucklin}, \bibinfo{person}{E Austin~Leone}, {and} \bibinfo{person}{Jenn Idema}.} \bibinfo{year}{2018}\natexlab{}.
\newblock \showarticletitle{Towards a Definition of Representational Competence}.
\newblock In \bibinfo{booktitle}{\emph{Towards a Framework for Representational Competence in Science Education}}. \bibinfo{publisher}{Springer}, \bibinfo{address}{Berlin, Germany}, \bibinfo{pages}{3--11}.
\newblock
\href{https://doi.org/10.1007/978-3-319-89945-9_1}{doi:\nolinkurl{10.1007/978-3-319-89945-9_1}}


\bibitem[D{\'e}tienne(1990)]%
        {detienne1990program}
\bibfield{author}{\bibinfo{person}{Fran{\c{c}}oise D{\'e}tienne}.} \bibinfo{year}{1990}\natexlab{}.
\newblock \showarticletitle{Program Understanding and Knowledge Organization: The Influence of Acquired Schemata}.
\newblock In \bibinfo{booktitle}{\emph{Cognitive Ergonomics: Understanding, Learning and Designing Human-Computer Interaction}}. \bibinfo{publisher}{Elsevier}, \bibinfo{address}{Amsterdam, Netherlands}, \bibinfo{pages}{245--256}.
\newblock
\href{https://doi.org/10.1016/b978-0-12-248290-8.50021-2}{doi:\nolinkurl{10.1016/b978-0-12-248290-8.50021-2}}


\bibitem[Duran et~al\mbox{.}(2022)]%
        {duran2022cognitive}
\bibfield{author}{\bibinfo{person}{Rodrigo Duran}, \bibinfo{person}{Albina Zavgorodniaia}, {and} \bibinfo{person}{Juha Sorva}.} \bibinfo{year}{2022}\natexlab{}.
\newblock \showarticletitle{Cognitive Load Theory in Computing Education Research: A Review}.
\newblock \bibinfo{journal}{\emph{ACM Transactions on Computing Education (TOCE)}} \bibinfo{volume}{22}, \bibinfo{number}{4} (\bibinfo{year}{2022}), \bibinfo{pages}{1--27}.
\newblock
\href{https://doi.org/10.1145/3483843}{doi:\nolinkurl{10.1145/3483843}}


\bibitem[Eckerdal and Thun{\'e}(2005)]%
        {eckerdal2005novice}
\bibfield{author}{\bibinfo{person}{Anna Eckerdal} {and} \bibinfo{person}{Michael Thun{\'e}}.} \bibinfo{year}{2005}\natexlab{}.
\newblock \showarticletitle{Novice Java Programmers' Conceptions Of" Object" And" Class", and Variation Theory}.
\newblock \bibinfo{journal}{\emph{ACM SIGCSE Bulletin}} \bibinfo{volume}{37}, \bibinfo{number}{3} (\bibinfo{year}{2005}), \bibinfo{pages}{89--93}.
\newblock
\href{https://doi.org/10.1145/1151954.1067473}{doi:\nolinkurl{10.1145/1151954.1067473}}


\bibitem[Eckert et~al\mbox{.}(2022)]%
        {eckert2022loops}
\bibfield{author}{\bibinfo{person}{Dimitri Eckert}, \bibinfo{person}{Dion Timmermann}, {and} \bibinfo{person}{Christian Kautz}.} \bibinfo{year}{2022}\natexlab{}.
\newblock \showarticletitle{Student Misconceptions About Loops in Introductory Programming Courses and the Influence of Representations}. In \bibinfo{booktitle}{\emph{2022 IEEE Frontiers in Education Conference (FIE)}}. \bibinfo{publisher}{IEEE}, \bibinfo{address}{Piscataway, NJ, USA}, \bibinfo{pages}{1--5}.
\newblock
\href{https://doi.org/10.1109/fie56618.2022.9962545}{doi:\nolinkurl{10.1109/fie56618.2022.9962545}}


\bibitem[Fincher et~al\mbox{.}(2020)]%
        {fincher2020notional}
\bibfield{author}{\bibinfo{person}{Sally Fincher}, \bibinfo{person}{Johan Jeuring}, \bibinfo{person}{Craig~S Miller}, \bibinfo{person}{Peter Donaldson}, \bibinfo{person}{Benedict Du~Boulay}, \bibinfo{person}{Matthias Hauswirth}, \bibinfo{person}{Arto Hellas}, \bibinfo{person}{Felienne Hermans}, \bibinfo{person}{Colleen Lewis}, \bibinfo{person}{Andreas M{\"u}hling}, {et~al\mbox{.}}} \bibinfo{year}{2020}\natexlab{}.
\newblock \showarticletitle{Notional Machines in Computing Education: The Education of Attention}.
\newblock In \bibinfo{booktitle}{\emph{Proceedings of the Working Group Reports on Innovation and Technology in Computer Science Education}}. \bibinfo{publisher}{ACM}, \bibinfo{address}{New York, NY, USA}, \bibinfo{pages}{21--50}.
\newblock


\bibitem[Fyfe et~al\mbox{.}(2014)]%
        {fyfe2014concreteness}
\bibfield{author}{\bibinfo{person}{Emily~R Fyfe}, \bibinfo{person}{Nicole~M McNeil}, \bibinfo{person}{Ji~Y Son}, {and} \bibinfo{person}{Robert~L Goldstone}.} \bibinfo{year}{2014}\natexlab{}.
\newblock \showarticletitle{Concreteness Fading in Mathematics and Science Instruction: A Systematic Review}.
\newblock \bibinfo{journal}{\emph{Educational Psychology Review}}  \bibinfo{volume}{26} (\bibinfo{year}{2014}), \bibinfo{pages}{9--25}.
\newblock
\href{https://doi.org/10.1007/s10648-014-9249-3}{doi:\nolinkurl{10.1007/s10648-014-9249-3}}


\bibitem[Gaver et~al\mbox{.}(2003)]%
        {gaver2003ambiguity}
\bibfield{author}{\bibinfo{person}{William~W Gaver}, \bibinfo{person}{Jacob Beaver}, {and} \bibinfo{person}{Steve Benford}.} \bibinfo{year}{2003}\natexlab{}.
\newblock \showarticletitle{Ambiguity as a Resource for Design}. In \bibinfo{booktitle}{\emph{Proceedings of the SIGCHI Conference on Human Factors in Computing Systems}}. \bibinfo{publisher}{ACM}, \bibinfo{address}{New York, NY, USA}, \bibinfo{pages}{233--240}.
\newblock
\href{https://doi.org/10.1145/642611.642653}{doi:\nolinkurl{10.1145/642611.642653}}


\bibitem[Gee(2000)]%
        {gee2000chapter}
\bibfield{author}{\bibinfo{person}{James~Paul Gee}.} \bibinfo{year}{2000}\natexlab{}.
\newblock \showarticletitle{Chapter 3: Identity as an Analytic Lens for Research in Education}.
\newblock \bibinfo{journal}{\emph{Review of Research in Education}} \bibinfo{volume}{25}, \bibinfo{number}{1} (\bibinfo{year}{2000}), \bibinfo{pages}{99--125}.
\newblock
\href{https://doi.org/10.3102/0091732x025001099}{doi:\nolinkurl{10.3102/0091732x025001099}}


\bibitem[Greca and Moreira(2000)]%
        {greca2000mental}
\bibfield{author}{\bibinfo{person}{Ileana~Maria Greca} {and} \bibinfo{person}{Marco~Antonio Moreira}.} \bibinfo{year}{2000}\natexlab{}.
\newblock \showarticletitle{Mental Models, Conceptual Models, and Modelling}.
\newblock \bibinfo{journal}{\emph{International Journal of Science Education}} \bibinfo{volume}{22}, \bibinfo{number}{1} (\bibinfo{year}{2000}), \bibinfo{pages}{1--11}.
\newblock
\href{https://doi.org/10.1080/095006900289976}{doi:\nolinkurl{10.1080/095006900289976}}


\bibitem[Guo(2013)]%
        {guo2013online}
\bibfield{author}{\bibinfo{person}{Philip~J Guo}.} \bibinfo{year}{2013}\natexlab{}.
\newblock \showarticletitle{Online Python Tutor: Embeddable Web-Based Program Visualization for CS Education}. In \bibinfo{booktitle}{\emph{Proceeding of the 44th ACM Technical Symposium on Computer Science Education}}. \bibinfo{publisher}{ACM}, \bibinfo{address}{New York, NY, USA}, \bibinfo{pages}{579--584}.
\newblock


\bibitem[Gupta and Rybarczyk(2023)]%
        {gupta2023improving}
\bibfield{author}{\bibinfo{person}{Ankur Gupta} {and} \bibinfo{person}{Ryan Rybarczyk}.} \bibinfo{year}{2023}\natexlab{}.
\newblock \showarticletitle{Improving Long Term Performance Using Visualized Scope Tracing: A 10-Year Study}. In \bibinfo{booktitle}{\emph{Proceedings of the 54th ACM Technical Symposium on Computer Science Education v. 1}}. \bibinfo{publisher}{ACM}, \bibinfo{address}{New York, NY, USA}, \bibinfo{pages}{137--143}.
\newblock
\href{https://doi.org/10.1145/3545945.3569748}{doi:\nolinkurl{10.1145/3545945.3569748}}


\bibitem[Hansen et~al\mbox{.}(2002)]%
        {hansen2002designing}
\bibfield{author}{\bibinfo{person}{Steven Hansen}, \bibinfo{person}{N~Hari Narayanan}, {and} \bibinfo{person}{Mary Hegarty}.} \bibinfo{year}{2002}\natexlab{}.
\newblock \showarticletitle{Designing Educationally Effective Algorithm Visualizations}.
\newblock \bibinfo{journal}{\emph{Journal of Visual Languages \& Computing}} \bibinfo{volume}{13}, \bibinfo{number}{3} (\bibinfo{year}{2002}), \bibinfo{pages}{291--317}.
\newblock
\href{https://doi.org/10.1006/jvlc.2002.0236}{doi:\nolinkurl{10.1006/jvlc.2002.0236}}


\bibitem[Hassan et~al\mbox{.}(2024)]%
        {hassan2024evaluating}
\bibfield{author}{\bibinfo{person}{Mohammed Hassan}, \bibinfo{person}{Grace Zeng}, {and} \bibinfo{person}{Craig Zilles}.} \bibinfo{year}{2024}\natexlab{}.
\newblock \showarticletitle{Evaluating How Novices Utilize Debuggers and Code Execution to Understand Code}. In \bibinfo{booktitle}{\emph{Proceedings of the 2024 ACM Conference on International Computing Education Research-Volume 1}}. \bibinfo{publisher}{ACM}, \bibinfo{address}{New York, NY, USA}, \bibinfo{pages}{65--83}.
\newblock
\href{https://doi.org/10.1145/3632620.3671126}{doi:\nolinkurl{10.1145/3632620.3671126}}


\bibitem[Hayatpur et~al\mbox{.}(2023)]%
        {hayatpur2023crosscode}
\bibfield{author}{\bibinfo{person}{Devamardeep Hayatpur}, \bibinfo{person}{Daniel Wigdor}, {and} \bibinfo{person}{Haijun Xia}.} \bibinfo{year}{2023}\natexlab{}.
\newblock \showarticletitle{Crosscode: Multi-Level Visualization of Program Execution}. In \bibinfo{booktitle}{\emph{Proceedings of the 2023 CHI Conference on Human Factors in Computing Systems}}. \bibinfo{publisher}{ACM}, \bibinfo{address}{New York, NY, USA}, \bibinfo{pages}{1--13}.
\newblock
\href{https://doi.org/10.1145/3544548.3581390}{doi:\nolinkurl{10.1145/3544548.3581390}}


\bibitem[Heer and Shneiderman(2012)]%
        {heer2012interactive}
\bibfield{author}{\bibinfo{person}{Jeffrey Heer} {and} \bibinfo{person}{Ben Shneiderman}.} \bibinfo{year}{2012}\natexlab{}.
\newblock \showarticletitle{Interactive Dynamics for Visual Analysis: A Taxonomy of Tools That Support the Fluent and Flexible Use of Visualizations}.
\newblock \bibinfo{journal}{\emph{Queue}} \bibinfo{volume}{10}, \bibinfo{number}{2} (\bibinfo{year}{2012}), \bibinfo{pages}{30--55}.
\newblock


\bibitem[Hegarty et~al\mbox{.}(2010)]%
        {hegarty2010thinking}
\bibfield{author}{\bibinfo{person}{Mary Hegarty}, \bibinfo{person}{Matt~S Canham}, {and} \bibinfo{person}{Sara~I Fabrikant}.} \bibinfo{year}{2010}\natexlab{}.
\newblock \showarticletitle{Thinking About the Weather: How Display Salience and Knowledge Affect Performance in a Graphic Inference Task.}
\newblock \bibinfo{journal}{\emph{Journal of Experimental Psychology: Learning, Memory, and Cognition}} \bibinfo{volume}{36}, \bibinfo{number}{1} (\bibinfo{year}{2010}), \bibinfo{pages}{37}.
\newblock
\href{https://doi.org/10.1037/a0017683}{doi:\nolinkurl{10.1037/a0017683}}


\bibitem[Heinonen et~al\mbox{.}(2023)]%
        {heinonen2023synthesizing}
\bibfield{author}{\bibinfo{person}{Ava Heinonen}, \bibinfo{person}{Bettina Lehtel{\"a}}, \bibinfo{person}{Arto Hellas}, {and} \bibinfo{person}{Fabian Fagerholm}.} \bibinfo{year}{2023}\natexlab{}.
\newblock \showarticletitle{Synthesizing Research on Programmers’ Mental Models of Programs, Tasks and Concepts—A Systematic Literature Review}.
\newblock \bibinfo{journal}{\emph{Information and Software Technology}}  \bibinfo{volume}{164} (\bibinfo{year}{2023}), \bibinfo{pages}{107300}.
\newblock
\href{https://doi.org/10.1016/j.infsof.2023.107300}{doi:\nolinkurl{10.1016/j.infsof.2023.107300}}


\bibitem[Hermans et~al\mbox{.}(2018)]%
        {hermans2018thinking}
\bibfield{author}{\bibinfo{person}{Felienne Hermans}, \bibinfo{person}{Alaaeddin Swidan}, \bibinfo{person}{Efthimia Aivaloglou}, {and} \bibinfo{person}{Marileen Smit}.} \bibinfo{year}{2018}\natexlab{}.
\newblock \showarticletitle{Thinking Out of the Box: Comparing Metaphors for Variables in Programming Education}. In \bibinfo{booktitle}{\emph{Proceedings of the 13th Workshop in Primary and Secondary Computing Education}}. \bibinfo{publisher}{ACM}, \bibinfo{address}{New York, NY, USA}, \bibinfo{pages}{1--8}.
\newblock


\bibitem[Hertz(2010)]%
        {hertz2010cs1}
\bibfield{author}{\bibinfo{person}{Matthew Hertz}.} \bibinfo{year}{2010}\natexlab{}.
\newblock \showarticletitle{What Do" CS1" And" CS2" Mean? Investigating Differences in the Early Courses}. In \bibinfo{booktitle}{\emph{Proceedings of the 41st ACM Technical Symposium on Computer Science Education}}. \bibinfo{publisher}{ACM}, \bibinfo{address}{New York, NY, USA}, \bibinfo{pages}{199--203}.
\newblock


\bibitem[Hill et~al\mbox{.}(2015)]%
        {hill2015online}
\bibfield{author}{\bibinfo{person}{M Hill}, \bibinfo{person}{MD Sharma}, {and} \bibinfo{person}{Helen Johnston}.} \bibinfo{year}{2015}\natexlab{}.
\newblock \showarticletitle{How Online Learning Modules Can Improve the Representational Fluency and Conceptual Understanding of University Physics Students}.
\newblock \bibinfo{journal}{\emph{European Journal of Physics}} \bibinfo{volume}{36}, \bibinfo{number}{4} (\bibinfo{year}{2015}), \bibinfo{pages}{045019}.
\newblock
\href{https://doi.org/10.1088/0143-0807/36/4/045019}{doi:\nolinkurl{10.1088/0143-0807/36/4/045019}}


\bibitem[Hundhausen et~al\mbox{.}(2002)]%
        {hundhausen2002meta}
\bibfield{author}{\bibinfo{person}{Christopher~D Hundhausen}, \bibinfo{person}{Sarah~A Douglas}, {and} \bibinfo{person}{John~T Stasko}.} \bibinfo{year}{2002}\natexlab{}.
\newblock \showarticletitle{A Meta-Study of Algorithm Visualization Effectiveness}.
\newblock \bibinfo{journal}{\emph{Journal of Visual Languages \& Computing}} \bibinfo{volume}{13}, \bibinfo{number}{3} (\bibinfo{year}{2002}), \bibinfo{pages}{259--290}.
\newblock
\href{https://doi.org/10.1006/jvlc.2002.0237}{doi:\nolinkurl{10.1006/jvlc.2002.0237}}


\bibitem[Isohanni and Järvinen(2014)]%
        {isohanni_are_2014}
\bibfield{author}{\bibinfo{person}{Essi Isohanni} {and} \bibinfo{person}{Hannu-Matti Järvinen}.} \bibinfo{year}{2014}\natexlab{}.
\newblock \showarticletitle{Are Visualization Tools Used in Programming Education?: By Whom, How, Why, and Why Not?}. In \bibinfo{booktitle}{\emph{Proceedings of the 14th {Koli} {Calling} {International} {Conference} on {Computing} {Education} {Research}}}. \bibinfo{publisher}{ACM}, \bibinfo{address}{Koli Finland}, \bibinfo{pages}{35--40}.
\newblock
\showISBNx{978-1-4503-3065-7}
\href{https://doi.org/10.1145/2674683.2674688}{doi:\nolinkurl{10.1145/2674683.2674688}}


\bibitem[Isohanni and Knobelsdorf(2013)]%
        {isohanni2013students}
\bibfield{author}{\bibinfo{person}{Essi Isohanni} {and} \bibinfo{person}{Maria Knobelsdorf}.} \bibinfo{year}{2013}\natexlab{}.
\newblock \bibinfo{booktitle}{\emph{Students' Engagement with the Visualization Tool VIP in Light of Activity Theory}}.
\newblock \bibinfo{publisher}{Tampere University of Technology. Department of Pervasive Computing}, \bibinfo{address}{Tampere, Finland}.
\newblock
\newblock
\shownote{Contribution: organisation=tie,FACT1=1<br/>Portfolio EDEND: 2015-02-27}.


\bibitem[Johnson-Laird(1986)]%
        {johnson1983mental}
\bibfield{author}{\bibinfo{person}{P.~N. Johnson-Laird}.} \bibinfo{year}{1986}\natexlab{}.
\newblock \bibinfo{booktitle}{\emph{Mental models: Towards a Cognitive Science of Language, Inference, and Consciousness}}.
\newblock \bibinfo{publisher}{Harvard University Press}, \bibinfo{address}{USA}.
\newblock
\showISBNx{0674568826}


\bibitem[Just and Carpenter(1976)]%
        {just1976eye}
\bibfield{author}{\bibinfo{person}{Marcel~Adam Just} {and} \bibinfo{person}{Patricia~A Carpenter}.} \bibinfo{year}{1976}\natexlab{}.
\newblock \showarticletitle{Eye Fixations and Cognitive Processes}.
\newblock \bibinfo{journal}{\emph{Cognitive Psychology}} \bibinfo{volume}{8}, \bibinfo{number}{4} (\bibinfo{year}{1976}), \bibinfo{pages}{441--480}.
\newblock
\href{https://doi.org/10.1016/0010-0285(76)90015-3}{doi:\nolinkurl{10.1016/0010-0285(76)90015-3}}


\bibitem[Kalyuga(2007)]%
        {kalyuga2007expertise}
\bibfield{author}{\bibinfo{person}{Slava Kalyuga}.} \bibinfo{year}{2007}\natexlab{}.
\newblock \showarticletitle{Expertise Reversal Effect and Its Implications for Learner-Tailored Instruction}.
\newblock \bibinfo{journal}{\emph{Educational Psychology Review}} \bibinfo{volume}{19}, \bibinfo{number}{4} (\bibinfo{year}{2007}), \bibinfo{pages}{509--539}.
\newblock
\href{https://doi.org/10.1007/s10648-007-9054-3}{doi:\nolinkurl{10.1007/s10648-007-9054-3}}


\bibitem[Kang and Guo(2017)]%
        {kang2017omnicode}
\bibfield{author}{\bibinfo{person}{Hyeonsu Kang} {and} \bibinfo{person}{Philip~J Guo}.} \bibinfo{year}{2017}\natexlab{}.
\newblock \showarticletitle{Omnicode: A Novice-Oriented Live Programming Environment With Always-on Run-Time Value Visualizations}. In \bibinfo{booktitle}{\emph{Proceedings of the 30th Annual ACM Symposium on User Interface Software and Technology}}. \bibinfo{publisher}{ACM}, \bibinfo{address}{New York, NY, USA}, \bibinfo{pages}{737--745}.
\newblock


\bibitem[Karnalim and Ayub(2017a)]%
        {karnalim2017effectiveness}
\bibfield{author}{\bibinfo{person}{Oscar Karnalim} {and} \bibinfo{person}{Mewati Ayub}.} \bibinfo{year}{2017}\natexlab{a}.
\newblock \showarticletitle{The Effectiveness of a Program Visualization Tool on Introductory Programming: A Case Study With PythonTutor}.
\newblock \bibinfo{journal}{\emph{CommIT (Communication and Information Technology) Journal}} \bibinfo{volume}{11}, \bibinfo{number}{2} (\bibinfo{year}{2017}), \bibinfo{pages}{67--76}.
\newblock
\href{https://doi.org/10.21512/commit.v11i2.3704}{doi:\nolinkurl{10.21512/commit.v11i2.3704}}


\bibitem[Karnalim and Ayub(2017b)]%
        {karnalim2017use}
\bibfield{author}{\bibinfo{person}{Oscar Karnalim} {and} \bibinfo{person}{Mewati Ayub}.} \bibinfo{year}{2017}\natexlab{b}.
\newblock \showarticletitle{The Use of Python Tutor on Programming Laboratory Session: Student Perspectives}.
\newblock \bibinfo{journal}{\emph{Kinetik: Game Technology, Information System, Computer Network, Computing, Electronics, and Control}} \bibinfo{volume}{2}, \bibinfo{number}{4} (\bibinfo{date}{Oct.} \bibinfo{year}{2017}), \bibinfo{pages}{327 -- 336}.
\newblock
\href{https://doi.org/10.22219/kinetik.v2i4.442}{doi:\nolinkurl{10.22219/kinetik.v2i4.442}}


\bibitem[Kirk-Johnson et~al\mbox{.}(2019)]%
        {kirk2019perceiving}
\bibfield{author}{\bibinfo{person}{Afton Kirk-Johnson}, \bibinfo{person}{Brian~M Galla}, {and} \bibinfo{person}{Scott~H Fraundorf}.} \bibinfo{year}{2019}\natexlab{}.
\newblock \showarticletitle{Perceiving Effort as Poor Learning: The Misinterpreted-Effort Hypothesis of How Experienced Effort and Perceived Learning Relate to Study Strategy Choice}.
\newblock \bibinfo{journal}{\emph{Cognitive Psychology}}  \bibinfo{volume}{115} (\bibinfo{year}{2019}), \bibinfo{pages}{101237}.
\newblock
\href{https://doi.org/10.1016/j.cogpsych.2019.101237}{doi:\nolinkurl{10.1016/j.cogpsych.2019.101237}}


\bibitem[Klepsch and Seufert(2021)]%
        {klepsch2021making}
\bibfield{author}{\bibinfo{person}{Melina Klepsch} {and} \bibinfo{person}{Tina Seufert}.} \bibinfo{year}{2021}\natexlab{}.
\newblock \showarticletitle{Making an Effort Versus Experiencing Load}. In \bibinfo{booktitle}{\emph{Frontiers in Education}}, Vol.~\bibinfo{volume}{6}. Frontiers Media SA, \bibinfo{publisher}{Frontiers Media SA}, \bibinfo{address}{Lausanne, Switzerland}, \bibinfo{pages}{645284}.
\newblock
\href{https://doi.org/10.3389/feduc.2021.645284}{doi:\nolinkurl{10.3389/feduc.2021.645284}}


\bibitem[Knobelsdorf et~al\mbox{.}(2012)]%
        {knobelsdorf2012reasons}
\bibfield{author}{\bibinfo{person}{Maria Knobelsdorf}, \bibinfo{person}{Essi Isohanni}, {and} \bibinfo{person}{Josh Tenenberg}.} \bibinfo{year}{2012}\natexlab{}.
\newblock \showarticletitle{The Reasons Might Be Different: Why Students and Teachers Do Not Use Visualization Tools}. In \bibinfo{booktitle}{\emph{Proceedings of the 12th Koli Calling International Conference on Computing Education Research}}. \bibinfo{publisher}{ACM}, \bibinfo{address}{New York, NY, USA}, \bibinfo{pages}{1--10}.
\newblock


\bibitem[Ko and Myers(2009)]%
        {ko2009whyline}
\bibfield{author}{\bibinfo{person}{Amy~J Ko} {and} \bibinfo{person}{Brad~A Myers}.} \bibinfo{year}{2009}\natexlab{}.
\newblock \showarticletitle{Finding Causes of Program Output With the Java Whyline}. In \bibinfo{booktitle}{\emph{Proceedings of the SIGCHI Conference on Human Factors in Computing Systems}}. \bibinfo{publisher}{ACM}, \bibinfo{address}{New York, NY, USA}, \bibinfo{pages}{1569--1578}.
\newblock
\href{https://doi.org/10.1145/1518701.1518942}{doi:\nolinkurl{10.1145/1518701.1518942}}


\bibitem[Lerner(2020)]%
        {lerner2020projection}
\bibfield{author}{\bibinfo{person}{Sorin Lerner}.} \bibinfo{year}{2020}\natexlab{}.
\newblock \showarticletitle{Projection Boxes: On-the-Fly Reconfigurable Visualization for Live Programming}. In \bibinfo{booktitle}{\emph{Proceedings of the 2020 CHI Conference on Human Factors in Computing Systems}}. \bibinfo{publisher}{ACM}, \bibinfo{address}{New York, NY, USA}, \bibinfo{pages}{1--7}.
\newblock
\href{https://doi.org/10.1145/3313831.3376494}{doi:\nolinkurl{10.1145/3313831.3376494}}


\bibitem[Levy and Feitelson(2019)]%
        {levy2019understanding}
\bibfield{author}{\bibinfo{person}{Omer Levy} {and} \bibinfo{person}{Dror~G Feitelson}.} \bibinfo{year}{2019}\natexlab{}.
\newblock \showarticletitle{Understanding Large-Scale Software--a Hierarchical View}. In \bibinfo{booktitle}{\emph{2019 IEEE/ACM 27th International Conference on Program Comprehension (ICPC)}}. \bibinfo{publisher}{IEEE}, \bibinfo{address}{Piscataway, NJ, USA}, \bibinfo{pages}{283--293}.
\newblock
\href{https://doi.org/10.1109/icpc.2019.00047}{doi:\nolinkurl{10.1109/icpc.2019.00047}}


\bibitem[Ma'ayan et~al\mbox{.}(2020)]%
        {ma2020domain}
\bibfield{author}{\bibinfo{person}{Dor Ma'ayan}, \bibinfo{person}{Wode Ni}, \bibinfo{person}{Katherine Ye}, \bibinfo{person}{Chinmay Kulkarni}, {and} \bibinfo{person}{Joshua Sunshine}.} \bibinfo{year}{2020}\natexlab{}.
\newblock \showarticletitle{How Domain Experts Create Conceptual Diagrams and Implications for Tool Design}. In \bibinfo{booktitle}{\emph{Proceedings of the 2020 CHI Conference on Human Factors in Computing Systems}}. \bibinfo{publisher}{ACM}, \bibinfo{address}{New York, NY, USA}, \bibinfo{pages}{1--14}.
\newblock
\href{https://doi.org/10.1145/3313831.3376253}{doi:\nolinkurl{10.1145/3313831.3376253}}


\bibitem[Mahanama et~al\mbox{.}(2022)]%
        {mahanama2022eye}
\bibfield{author}{\bibinfo{person}{Bhanuka Mahanama}, \bibinfo{person}{Yasith Jayawardana}, \bibinfo{person}{Sundararaman Rengarajan}, \bibinfo{person}{Gavindya Jayawardena}, \bibinfo{person}{Leanne Chukoskie}, \bibinfo{person}{Joseph Snider}, {and} \bibinfo{person}{Sampath Jayarathna}.} \bibinfo{year}{2022}\natexlab{}.
\newblock \showarticletitle{Eye Movement and Pupil Measures: A Review}.
\newblock \bibinfo{journal}{\emph{Frontiers in Computer Science}}  \bibinfo{volume}{3} (\bibinfo{year}{2022}), \bibinfo{pages}{733531}.
\newblock
\href{https://doi.org/10.3389/fcomp.2021.733531}{doi:\nolinkurl{10.3389/fcomp.2021.733531}}


\bibitem[Maton et~al\mbox{.}(2015)]%
        {maton2015knowledge}
\bibfield{author}{\bibinfo{person}{Karl Maton}, \bibinfo{person}{Susan Hood}, {and} \bibinfo{person}{Suellen Shay}.} \bibinfo{year}{2015}\natexlab{}.
\newblock \bibinfo{booktitle}{\emph{Knowledge-Building: Educational Studies in Legitimation Code Theory}}.
\newblock \bibinfo{publisher}{Routledge}, \bibinfo{address}{London, UK}.
\newblock
\href{https://doi.org/10.1080/14703297.2016.1231751}{doi:\nolinkurl{10.1080/14703297.2016.1231751}}


\bibitem[Mayer et~al\mbox{.}(2008)]%
        {mayer2008increased}
\bibfield{author}{\bibinfo{person}{Richard~E Mayer}, \bibinfo{person}{Emily Griffith}, \bibinfo{person}{Ilana~TN Jurkowitz}, {and} \bibinfo{person}{Daniel Rothman}.} \bibinfo{year}{2008}\natexlab{}.
\newblock \showarticletitle{Increased Interestingness of Extraneous Details in a Multimedia Science Presentation Leads to Decreased Learning.}
\newblock \bibinfo{journal}{\emph{Journal of Experimental Psychology: Applied}} \bibinfo{volume}{14}, \bibinfo{number}{4} (\bibinfo{year}{2008}), \bibinfo{pages}{329}.
\newblock
\href{https://doi.org/10.1037/a0013835}{doi:\nolinkurl{10.1037/a0013835}}


\bibitem[Mnguni(2014)]%
        {mnguni2014theoretical}
\bibfield{author}{\bibinfo{person}{Lindelani~E Mnguni}.} \bibinfo{year}{2014}\natexlab{}.
\newblock \showarticletitle{The Theoretical Cognitive Process of Visualization for Science Education}.
\newblock \bibinfo{journal}{\emph{SpringerPlus}}  \bibinfo{volume}{3} (\bibinfo{year}{2014}), \bibinfo{pages}{1--9}.
\newblock
\href{https://doi.org/10.1186/2193-1801-3-184}{doi:\nolinkurl{10.1186/2193-1801-3-184}}


\bibitem[Moraz{\'a}n(2020)]%
        {morazan2020make}
\bibfield{author}{\bibinfo{person}{Marco~T Moraz{\'a}n}.} \bibinfo{year}{2020}\natexlab{}.
\newblock \showarticletitle{How to Make While Loops Iterative: An Introduction for First-Year CS Students}. In \bibinfo{booktitle}{\emph{Proceedings of the 9th Computer Science Education Research Conference}}. \bibinfo{publisher}{ACM}, \bibinfo{address}{New York, NY, USA}, \bibinfo{pages}{1--12}.
\newblock


\bibitem[Munzner(2014)]%
        {munzner2014visualization}
\bibfield{author}{\bibinfo{person}{Tamara Munzner}.} \bibinfo{year}{2014}\natexlab{}.
\newblock \bibinfo{booktitle}{\emph{Visualization Analysis and Design}}.
\newblock \bibinfo{publisher}{CRC Press}, \bibinfo{address}{Boca Raton, FL, USA}.
\newblock
\href{https://doi.org/10.1201/b17511}{doi:\nolinkurl{10.1201/b17511}}


\bibitem[Naps et~al\mbox{.}(2002)]%
        {naps2002exploring}
\bibfield{author}{\bibinfo{person}{Thomas~L Naps}, \bibinfo{person}{Guido R{\"o}{\ss}ling}, \bibinfo{person}{Vicki Almstrum}, \bibinfo{person}{Wanda Dann}, \bibinfo{person}{Rudolf Fleischer}, \bibinfo{person}{Chris Hundhausen}, \bibinfo{person}{Ari Korhonen}, \bibinfo{person}{Lauri Malmi}, \bibinfo{person}{Myles McNally}, \bibinfo{person}{Susan Rodger}, {et~al\mbox{.}}} \bibinfo{year}{2002}\natexlab{}.
\newblock \showarticletitle{{Exploring the Role of Visualization and Engagement in Computer Science Education }}.
\newblock In \bibinfo{booktitle}{\emph{Working Group Reports From ITiCSE on Innovation and Technology in Computer Science Education}}. \bibinfo{publisher}{ACM}, \bibinfo{address}{New York, NY, USA}, \bibinfo{pages}{131--152}.
\newblock
\href{https://doi.org/10.1145/960568.782998}{doi:\nolinkurl{10.1145/960568.782998}}


\bibitem[Nasir and Hand(2006)]%
        {nasir2006exploring}
\bibfield{author}{\bibinfo{person}{Na’ilah~Suad Nasir} {and} \bibinfo{person}{Victoria~M Hand}.} \bibinfo{year}{2006}\natexlab{}.
\newblock \showarticletitle{Exploring Sociocultural Perspectives on Race, Culture, and Learning}.
\newblock \bibinfo{journal}{\emph{Review of Educational Research}} \bibinfo{volume}{76}, \bibinfo{number}{4} (\bibinfo{year}{2006}), \bibinfo{pages}{449--475}.
\newblock
\href{https://doi.org/10.3102/00346543076004449}{doi:\nolinkurl{10.3102/00346543076004449}}


\bibitem[Niehorster et~al\mbox{.}(2020)]%
        {niehorster2020characterizing}
\bibfield{author}{\bibinfo{person}{Diederick~C Niehorster}, \bibinfo{person}{Raimondas Zemblys}, \bibinfo{person}{Tanya Beelders}, {and} \bibinfo{person}{Kenneth Holmqvist}.} \bibinfo{year}{2020}\natexlab{}.
\newblock \showarticletitle{Characterizing Gaze Position Signals and Synthesizing Noise During Fixations in Eye-Tracking Data}.
\newblock \bibinfo{journal}{\emph{Behavior Research Methods}} \bibinfo{volume}{52}, \bibinfo{number}{6} (\bibinfo{year}{2020}), \bibinfo{pages}{2515--2534}.
\newblock
\href{https://doi.org/10.3758/s13428-020-01400-9}{doi:\nolinkurl{10.3758/s13428-020-01400-9}}


\bibitem[Paas and Ayres(2014)]%
        {paas2014cognitive}
\bibfield{author}{\bibinfo{person}{Fred Paas} {and} \bibinfo{person}{Paul Ayres}.} \bibinfo{year}{2014}\natexlab{}.
\newblock \showarticletitle{Cognitive Load Theory: A Broader View on the Role of Memory in Learning and Education}.
\newblock \bibinfo{journal}{\emph{Educational Psychology Review}}  \bibinfo{volume}{26} (\bibinfo{year}{2014}), \bibinfo{pages}{191--195}.
\newblock
\href{https://doi.org/10.1007/s10648-014-9263-5}{doi:\nolinkurl{10.1007/s10648-014-9263-5}}


\bibitem[Paas et~al\mbox{.}(2003)]%
        {paas2003cognitive}
\bibfield{author}{\bibinfo{person}{Fred Paas}, \bibinfo{person}{Alexander Renkl}, {and} \bibinfo{person}{John Sweller}.} \bibinfo{year}{2003}\natexlab{}.
\newblock \showarticletitle{Cognitive Load Theory and Instructional Design: Recent Developments}.
\newblock \bibinfo{journal}{\emph{Educational Psychologist}} \bibinfo{volume}{38}, \bibinfo{number}{1} (\bibinfo{year}{2003}), \bibinfo{pages}{1--4}.
\newblock
\href{https://doi.org/10.1207/s15326985ep3801_1}{doi:\nolinkurl{10.1207/s15326985ep3801_1}}


\bibitem[Papoutsaki et~al\mbox{.}(2018)]%
        {papoutsaki2018eye}
\bibfield{author}{\bibinfo{person}{Alexandra Papoutsaki}, \bibinfo{person}{Aaron Gokaslan}, \bibinfo{person}{James Tompkin}, \bibinfo{person}{Yuze He}, {and} \bibinfo{person}{Jeff Huang}.} \bibinfo{year}{2018}\natexlab{}.
\newblock \showarticletitle{The Eye of the Typer: A Benchmark and Analysis of Gaze Behavior During Typing}. In \bibinfo{booktitle}{\emph{Proceedings of the 2018 ACM Symposium on Eye Tracking Research \& Applications}}. \bibinfo{publisher}{ACM}, \bibinfo{address}{New York, NY, USA}, \bibinfo{pages}{1--9}.
\newblock


\bibitem[Papoutsaki et~al\mbox{.}(2017)]%
        {papoutsaki2017searchgazer}
\bibfield{author}{\bibinfo{person}{Alexandra Papoutsaki}, \bibinfo{person}{James Laskey}, {and} \bibinfo{person}{Jeff Huang}.} \bibinfo{year}{2017}\natexlab{}.
\newblock \showarticletitle{Searchgazer: Webcam Eye Tracking for Remote Studies of Web Search}. In \bibinfo{booktitle}{\emph{Proceedings of the 2017 Conference on Conference Human Information Interaction and Retrieval}}. \bibinfo{publisher}{ACM}, \bibinfo{address}{New York, NY, USA}, \bibinfo{pages}{17--26}.
\newblock


\bibitem[Permatasari et~al\mbox{.}(2022)]%
        {permatasari2022chemistry}
\bibfield{author}{\bibinfo{person}{Margaretha~Bhrizda Permatasari}, \bibinfo{person}{Sri Rahayu}, {and} \bibinfo{person}{I~Wayan Dasna}.} \bibinfo{year}{2022}\natexlab{}.
\newblock \showarticletitle{Chemistry Learning Using Multiple Representations: A Systematic Literature Review.}
\newblock \bibinfo{journal}{\emph{Journal of Science Learning}} \bibinfo{volume}{5}, \bibinfo{number}{2} (\bibinfo{year}{2022}), \bibinfo{pages}{334--341}.
\newblock
\href{https://doi.org/10.17509/jsl.v5i2.42656}{doi:\nolinkurl{10.17509/jsl.v5i2.42656}}


\bibitem[Pollock et~al\mbox{.}(2020)]%
        {pollock2020essence}
\bibfield{author}{\bibinfo{person}{Josh Pollock}, \bibinfo{person}{Grace Oh}, \bibinfo{person}{Eunice Jun}, \bibinfo{person}{Philip~J Guo}, {and} \bibinfo{person}{Zachary Tatlock}.} \bibinfo{year}{2020}\natexlab{}.
\newblock \bibinfo{title}{The Essence of Program Semantics Visualizers: A Three-Axis Model}.
\newblock


\bibitem[Porter et~al\mbox{.}(2018)]%
        {porter2018developing}
\bibfield{author}{\bibinfo{person}{Leo Porter}, \bibinfo{person}{Daniel Zingaro}, \bibinfo{person}{Cynthia Lee}, \bibinfo{person}{Cynthia Taylor}, \bibinfo{person}{Kevin~C Webb}, {and} \bibinfo{person}{Michael Clancy}.} \bibinfo{year}{2018}\natexlab{}.
\newblock \showarticletitle{Developing Course-Level Learning Goals for Basic Data Structures in CS2}. In \bibinfo{booktitle}{\emph{Proceedings of the 49th ACM Technical Symposium on Computer Science Education}}. \bibinfo{publisher}{ACM}, \bibinfo{address}{New York, NY, USA}, \bibinfo{pages}{858--863}.
\newblock
\href{https://doi.org/10.1145/3159450.3159457}{doi:\nolinkurl{10.1145/3159450.3159457}}


\bibitem[Reber and Greifeneder(2017)]%
        {reber2017processing}
\bibfield{author}{\bibinfo{person}{Rolf Reber} {and} \bibinfo{person}{Rainer Greifeneder}.} \bibinfo{year}{2017}\natexlab{}.
\newblock \showarticletitle{Processing Fluency in Education: How Metacognitive Feelings Shape Learning, Belief Formation, and Affect}.
\newblock \bibinfo{journal}{\emph{Educational Psychologist}} \bibinfo{volume}{52}, \bibinfo{number}{2} (\bibinfo{year}{2017}), \bibinfo{pages}{84--103}.
\newblock
\href{https://doi.org/10.1080/00461520.2016.1258173}{doi:\nolinkurl{10.1080/00461520.2016.1258173}}


\bibitem[Roberts and Wright(2006)]%
        {roberts2006towards}
\bibfield{author}{\bibinfo{person}{Jonathan~C Roberts} {and} \bibinfo{person}{Michael~AE Wright}.} \bibinfo{year}{2006}\natexlab{}.
\newblock \showarticletitle{Towards Ubiquitous Brushing for Information Visualization}. In \bibinfo{booktitle}{\emph{Tenth International Conference on Information Visualisation (IV'06)}}. \bibinfo{publisher}{IEEE}, \bibinfo{address}{Piscataway, NJ, USA}, \bibinfo{pages}{151--156}.
\newblock
\href{https://doi.org/10.1109/iv.2006.113}{doi:\nolinkurl{10.1109/iv.2006.113}}


\bibitem[Sajaniemi et~al\mbox{.}(2008)]%
        {sajaniemi2008study}
\bibfield{author}{\bibinfo{person}{Jorma Sajaniemi}, \bibinfo{person}{Marja Kuittinen}, {and} \bibinfo{person}{Taina Tikansalo}.} \bibinfo{year}{2008}\natexlab{}.
\newblock \showarticletitle{A Study of the Development of Students' Visualizations of Program State During an Elementary Object-Oriented Programming Course}.
\newblock \bibinfo{journal}{\emph{Journal on Educational Resources in Computing (JERIC)}} \bibinfo{volume}{7}, \bibinfo{number}{4} (\bibinfo{year}{2008}), \bibinfo{pages}{1--31}.
\newblock
\href{https://doi.org/10.1145/1316450.1316453}{doi:\nolinkurl{10.1145/1316450.1316453}}


\bibitem[Sanford et~al\mbox{.}(2014)]%
        {sanford2014metaphors}
\bibfield{author}{\bibinfo{person}{Joseph~P Sanford}, \bibinfo{person}{Aaron Tietz}, \bibinfo{person}{Saad Farooq}, \bibinfo{person}{Samuel Guyer}, {and} \bibinfo{person}{R~Benjamin Shapiro}.} \bibinfo{year}{2014}\natexlab{}.
\newblock \showarticletitle{Metaphors We Teach By}. In \bibinfo{booktitle}{\emph{Proceedings of the 45th ACM Technical Symposium on Computer Science Education}}. \bibinfo{publisher}{ACM}, \bibinfo{address}{New York, NY, USA}, \bibinfo{pages}{585--590}.
\newblock
\href{https://doi.org/10.1075/msw.19021.har}{doi:\nolinkurl{10.1075/msw.19021.har}}


\bibitem[Scaife and Rogers(1996)]%
        {scaife1996external}
\bibfield{author}{\bibinfo{person}{Mike Scaife} {and} \bibinfo{person}{Yvonne Rogers}.} \bibinfo{year}{1996}\natexlab{}.
\newblock \showarticletitle{External Cognition: How Do Graphical Representations Work?}
\newblock \bibinfo{journal}{\emph{International Journal of Human-Computer Studies}} \bibinfo{volume}{45}, \bibinfo{number}{2} (\bibinfo{year}{1996}), \bibinfo{pages}{185--213}.
\newblock
\href{https://doi.org/10.1006/ijhc.1996.0048}{doi:\nolinkurl{10.1006/ijhc.1996.0048}}


\bibitem[Scheiter et~al\mbox{.}(2020)]%
        {scheiter2020effort}
\bibfield{author}{\bibinfo{person}{Katharina Scheiter}, \bibinfo{person}{Rakefet Ackerman}, {and} \bibinfo{person}{Vincent Hoogerheide}.} \bibinfo{year}{2020}\natexlab{}.
\newblock \showarticletitle{Looking at Mental Effort Appraisals Through a Metacognitive Lens: Are They Biased?}
\newblock \bibinfo{journal}{\emph{Educational Psychology Review}} \bibinfo{volume}{32}, \bibinfo{number}{4} (\bibinfo{year}{2020}), \bibinfo{pages}{1003--1027}.
\newblock
\href{https://doi.org/10.1007/s10648-020-09555-9}{doi:\nolinkurl{10.1007/s10648-020-09555-9}}


\bibitem[Schwonke et~al\mbox{.}(2009)]%
        {schwonke2009multiple}
\bibfield{author}{\bibinfo{person}{Rolf Schwonke}, \bibinfo{person}{Kirsten Berthold}, {and} \bibinfo{person}{Alexander Renkl}.} \bibinfo{year}{2009}\natexlab{}.
\newblock \showarticletitle{How Multiple External Representations Are Used and How They Can Be Made More Useful}.
\newblock \bibinfo{journal}{\emph{Applied Cognitive Psychology: The Official Journal of the Society for Applied Research in Memory and Cognition}} \bibinfo{volume}{23}, \bibinfo{number}{9} (\bibinfo{year}{2009}), \bibinfo{pages}{1227--1243}.
\newblock
\href{https://doi.org/10.1002/acp.1526}{doi:\nolinkurl{10.1002/acp.1526}}


\bibitem[Seel(2006)]%
        {seel2006mental}
\bibfield{author}{\bibinfo{person}{Norbert~M Seel}.} \bibinfo{year}{2006}\natexlab{}.
\newblock \showarticletitle{Mental Models in Learning Situations}.
\newblock In \bibinfo{booktitle}{\emph{Advances in Psychology}}. Vol.~\bibinfo{volume}{138}. \bibinfo{publisher}{Elsevier}, \bibinfo{address}{Amsterdam, Netherlands}, \bibinfo{pages}{85--107}.
\newblock
\href{https://doi.org/10.1016/s0166-4115(06)80028-2}{doi:\nolinkurl{10.1016/s0166-4115(06)80028-2}}


\bibitem[Seel(2017)]%
        {seel2017model}
\bibfield{author}{\bibinfo{person}{Norbert~M Seel}.} \bibinfo{year}{2017}\natexlab{}.
\newblock \showarticletitle{Model-Based Learning: A Synthesis of Theory and Research}.
\newblock \bibinfo{journal}{\emph{Educational Technology Research and Development}}  \bibinfo{volume}{65} (\bibinfo{year}{2017}), \bibinfo{pages}{931--966}.
\newblock
\href{https://doi.org/10.1007/s11423-016-9507-9}{doi:\nolinkurl{10.1007/s11423-016-9507-9}}


\bibitem[Shaffer et~al\mbox{.}(2010)]%
        {shaffer2010algorithm}
\bibfield{author}{\bibinfo{person}{Clifford~A Shaffer}, \bibinfo{person}{Matthew~L Cooper}, \bibinfo{person}{Alexander Joel~D Alon}, \bibinfo{person}{Monika Akbar}, \bibinfo{person}{Michael Stewart}, \bibinfo{person}{Sean Ponce}, {and} \bibinfo{person}{Stephen~H Edwards}.} \bibinfo{year}{2010}\natexlab{}.
\newblock \showarticletitle{Algorithm Visualization: The State of the Field}.
\newblock \bibinfo{journal}{\emph{ACM Transactions on Computing Education (TOCE)}} \bibinfo{volume}{10}, \bibinfo{number}{3} (\bibinfo{year}{2010}), \bibinfo{pages}{1--22}.
\newblock


\bibitem[Sibia et~al\mbox{.}(2025a)]%
        {sibia2025code}
\bibfield{author}{\bibinfo{person}{Naaz Sibia}, \bibinfo{person}{Valeria~Ramirez Osorio}, \bibinfo{person}{Jessica Wen}, \bibinfo{person}{Rutwa Engineer}, \bibinfo{person}{Angela~Zavaleta Bernuy}, \bibinfo{person}{Andrew Petersen}, \bibinfo{person}{Michael Liut}, {and} \bibinfo{person}{Carolina Nobre}.} \bibinfo{year}{2025}\natexlab{a}.
\newblock \bibinfo{title}{From Code to Concept: Evaluating Multiple Coordinated Views in Introductory Programming}.
\newblock
\urldef\tempurl%
\url{https://arxiv.org/abs/2509.26466}
\showURL{%
\tempurl}


\bibitem[Sibia et~al\mbox{.}(2025b)]%
        {sibia2025state}
\bibfield{author}{\bibinfo{person}{Naaz Sibia}, \bibinfo{person}{Jessica Wen}, \bibinfo{person}{Amber Richardson}, \bibinfo{person}{Yashika Jain}, \bibinfo{person}{Angela Zavaleta~Bernuy}, \bibinfo{person}{Bogdan Simion}, \bibinfo{person}{Andrew Petersen}, \bibinfo{person}{Carolina Nobre}, {and} \bibinfo{person}{Michael Liut}.} \bibinfo{year}{2025}\natexlab{b}.
\newblock \showarticletitle{From State to Structure: Towards Abstraction Support in CS2}. In \bibinfo{booktitle}{\emph{Proceedings of the 25th Koli Calling International Conference on Computing Education Research}}. \bibinfo{publisher}{ACM}, \bibinfo{address}{New York, NY, USA}, \bibinfo{pages}{1--12}.
\newblock
\href{https://doi.org/10.1145/3769994.3769998}{doi:\nolinkurl{10.1145/3769994.3769998}}


\bibitem[Sillito et~al\mbox{.}(2008)]%
        {sillito2008asking}
\bibfield{author}{\bibinfo{person}{Jonathan Sillito}, \bibinfo{person}{Gail~C Murphy}, {and} \bibinfo{person}{Kris De~Volder}.} \bibinfo{year}{2008}\natexlab{}.
\newblock \showarticletitle{Asking and Answering Questions During a Programming Change Task}.
\newblock \bibinfo{journal}{\emph{IEEE Transactions on Software Engineering}} \bibinfo{volume}{34}, \bibinfo{number}{4} (\bibinfo{year}{2008}), \bibinfo{pages}{434--451}.
\newblock
\href{https://doi.org/10.1109/tse.2008.26}{doi:\nolinkurl{10.1109/tse.2008.26}}


\bibitem[Sirki{\"a}(2018)]%
        {sirkia2018jsvee}
\bibfield{author}{\bibinfo{person}{Teemu Sirki{\"a}}.} \bibinfo{year}{2018}\natexlab{}.
\newblock \showarticletitle{Jsvee \& Kelmu: Creating and Tailoring Program Animations for Computing Education}.
\newblock \bibinfo{journal}{\emph{Journal of Software: Evolution and Process}} \bibinfo{volume}{30}, \bibinfo{number}{2} (\bibinfo{year}{2018}), \bibinfo{pages}{e1924}.
\newblock
\href{https://doi.org/10.1109/vissoft.2016.24}{doi:\nolinkurl{10.1109/vissoft.2016.24}}


\bibitem[Slim and Hartsuiker(2023)]%
        {slim2023moving}
\bibfield{author}{\bibinfo{person}{Mieke~Sarah Slim} {and} \bibinfo{person}{Robert~J Hartsuiker}.} \bibinfo{year}{2023}\natexlab{}.
\newblock \showarticletitle{Moving Visual World Experiments Online? A Web-Based Replication of Dijkgraaf, Hartsuiker, and Duyck (2017) Using PCIbex and WebGazer. Js}.
\newblock \bibinfo{journal}{\emph{Behavior Research Methods}} \bibinfo{volume}{55}, \bibinfo{number}{7} (\bibinfo{year}{2023}), \bibinfo{pages}{3786--3804}.
\newblock
\href{https://doi.org/10.3758/s13428-022-01989-z}{doi:\nolinkurl{10.3758/s13428-022-01989-z}}


\bibitem[Sorva(2013)]%
        {juha2013notional}
\bibfield{author}{\bibinfo{person}{Juha Sorva}.} \bibinfo{year}{2013}\natexlab{}.
\newblock \showarticletitle{Notional Machines and Introductory Programming Education}.
\newblock \bibinfo{journal}{\emph{ACM Transactions on Computing Education}} \bibinfo{volume}{13}, \bibinfo{number}{2}, Article \bibinfo{articleno}{8} (\bibinfo{date}{July} \bibinfo{year}{2013}), \bibinfo{numpages}{31}~pages.
\newblock
\href{https://doi.org/10.1145/2483710.2483713}{doi:\nolinkurl{10.1145/2483710.2483713}}


\bibitem[Sorva et~al\mbox{.}(2013)]%
        {sorva2013review}
\bibfield{author}{\bibinfo{person}{Juha Sorva}, \bibinfo{person}{Ville Karavirta}, {and} \bibinfo{person}{Lauri Malmi}.} \bibinfo{year}{2013}\natexlab{}.
\newblock \showarticletitle{A Review of Generic Program Visualization Systems for Introductory Programming Education}.
\newblock \bibinfo{journal}{\emph{ACM Transactions on Computing Education (TOCE)}} \bibinfo{volume}{13}, \bibinfo{number}{4} (\bibinfo{year}{2013}), \bibinfo{pages}{1--64}.
\newblock
\href{https://doi.org/10.1145/2490822}{doi:\nolinkurl{10.1145/2490822}}


\bibitem[Springer and Whittaker(2019)]%
        {springer2019progressive}
\bibfield{author}{\bibinfo{person}{Aaron Springer} {and} \bibinfo{person}{Steve Whittaker}.} \bibinfo{year}{2019}\natexlab{}.
\newblock \showarticletitle{Progressive Disclosure: Empirically Motivated Approaches to Designing Effective Transparency}. In \bibinfo{booktitle}{\emph{Proceedings of the 24th International Conference on Intelligent User Interfaces}}. \bibinfo{publisher}{ACM}, \bibinfo{address}{New York, NY, USA}, \bibinfo{pages}{107--120}.
\newblock


\bibitem[Suh and Moyer-Packenham(2007)]%
        {suh2007developing}
\bibfield{author}{\bibinfo{person}{Jennifer Suh} {and} \bibinfo{person}{Patricia Moyer-Packenham}.} \bibinfo{year}{2007}\natexlab{}.
\newblock \showarticletitle{Developing Students’ Representational Fluency Using Virtual and Physical Algebra Balances}.
\newblock \bibinfo{journal}{\emph{Journal of Computers in Mathematics and Science Teaching}} \bibinfo{volume}{26}, \bibinfo{number}{2} (\bibinfo{year}{2007}), \bibinfo{pages}{155--173}.
\newblock


\bibitem[Suh et~al\mbox{.}(2020)]%
        {suh2020coding}
\bibfield{author}{\bibinfo{person}{Sangho Suh}, \bibinfo{person}{Martinet Lee}, \bibinfo{person}{Gracie Xia}, {et~al\mbox{.}}} \bibinfo{year}{2020}\natexlab{}.
\newblock \showarticletitle{Coding Strip: A Pedagogical Tool for Teaching and Learning Programming Concepts Through Comics}. In \bibinfo{booktitle}{\emph{2020 IEEE Symposium on Visual Languages and Human-Centric Computing (VL/HCC)}}. \bibinfo{publisher}{IEEE}, \bibinfo{address}{Piscataway, NJ, USA}, \bibinfo{pages}{1--10}.
\newblock
\href{https://doi.org/10.1109/vl/hcc50065.2020.9127262}{doi:\nolinkurl{10.1109/vl/hcc50065.2020.9127262}}


\bibitem[Suh et~al\mbox{.}(2022)]%
        {suh2022codetoon}
\bibfield{author}{\bibinfo{person}{Sangho Suh}, \bibinfo{person}{Jian Zhao}, {and} \bibinfo{person}{Edith Law}.} \bibinfo{year}{2022}\natexlab{}.
\newblock \showarticletitle{Codetoon: Story Ideation, Auto Comic Generation, and Structure Mapping for Code-Driven Storytelling}. In \bibinfo{booktitle}{\emph{Proceedings of the 35th Annual ACM Symposium on User Interface Software and Technology}}. \bibinfo{publisher}{ACM}, \bibinfo{address}{New York, NY, USA}, \bibinfo{pages}{1--16}.
\newblock
\href{https://doi.org/10.1145/3526113.3545617}{doi:\nolinkurl{10.1145/3526113.3545617}}


\bibitem[Sweller(2011)]%
        {sweller2011cognitive}
\bibfield{author}{\bibinfo{person}{John Sweller}.} \bibinfo{year}{2011}\natexlab{}.
\newblock \showarticletitle{Cognitive Load Theory}.
\newblock In \bibinfo{booktitle}{\emph{Psychology of Learning and Motivation}}. Vol.~\bibinfo{volume}{55}. \bibinfo{publisher}{Elsevier}, \bibinfo{address}{Amsterdam, Netherlands}, \bibinfo{pages}{37--76}.
\newblock
\href{https://doi.org/10.4324/9781003334361-4}{doi:\nolinkurl{10.4324/9781003334361-4}}


\bibitem[Treagust et~al\mbox{.}(2017)]%
        {treagust2017multiple}
\bibfield{author}{\bibinfo{person}{David~F Treagust}, \bibinfo{person}{Reinders Duit}, {and} \bibinfo{person}{Hans~E Fischer}.} \bibinfo{year}{2017}\natexlab{}.
\newblock \bibinfo{booktitle}{\emph{Multiple Representations in Physics Education}}. Vol.~\bibinfo{volume}{10}.
\newblock \bibinfo{publisher}{Springer}, \bibinfo{address}{Berlin, Germany}.
\newblock
\href{https://doi.org/10.1007/978-3-319-58914-5}{doi:\nolinkurl{10.1007/978-3-319-58914-5}}


\bibitem[Tripathi(2008)]%
        {tripathi2008developing}
\bibfield{author}{\bibinfo{person}{Preety~N Tripathi}.} \bibinfo{year}{2008}\natexlab{}.
\newblock \showarticletitle{Developing Mathematical Understanding Through Multiple Representations}.
\newblock \bibinfo{journal}{\emph{Mathematics Teaching in the Middle School}} \bibinfo{volume}{13}, \bibinfo{number}{8} (\bibinfo{year}{2008}), \bibinfo{pages}{438--445}.
\newblock
\href{https://doi.org/10.5951/mtms.13.8.0438}{doi:\nolinkurl{10.5951/mtms.13.8.0438}}


\bibitem[Underwood et~al\mbox{.}(2004)]%
        {underwood2004inspecting}
\bibfield{author}{\bibinfo{person}{Geoffrey Underwood}, \bibinfo{person}{Lorraine Jebbett}, {and} \bibinfo{person}{Katharine Roberts}.} \bibinfo{year}{2004}\natexlab{}.
\newblock \showarticletitle{Inspecting Pictures for Information to Verify a Sentence: Eye Movements in General Encoding and in Focused Search}.
\newblock \bibinfo{journal}{\emph{Quarterly Journal of Experimental Psychology Section A}} \bibinfo{volume}{57}, \bibinfo{number}{1} (\bibinfo{year}{2004}), \bibinfo{pages}{165--182}.
\newblock
\href{https://doi.org/10.1080/02724980343000189}{doi:\nolinkurl{10.1080/02724980343000189}}


\bibitem[Vel{\'a}zquez-Iturbide and P{\'e}rez-Carrasco(2010)]%
        {velazquez2010infovis}
\bibfield{author}{\bibinfo{person}{J~{\'A}ngel Vel{\'a}zquez-Iturbide} {and} \bibinfo{person}{Antonio P{\'e}rez-Carrasco}.} \bibinfo{year}{2010}\natexlab{}.
\newblock \showarticletitle{InfoVis Interaction Techniques in Animation of Recursive Programs}.
\newblock \bibinfo{journal}{\emph{Algorithms}} \bibinfo{volume}{3}, \bibinfo{number}{1} (\bibinfo{year}{2010}), \bibinfo{pages}{76--91}.
\newblock
\href{https://doi.org/10.1201/b13124-12}{doi:\nolinkurl{10.1201/b13124-12}}


\bibitem[Vel{\'a}zquez-Iturbide and P{\'e}rez-Carrasco(2016)]%
        {velazquez2016srec}
\bibfield{author}{\bibinfo{person}{J~{\'A}ngel Vel{\'a}zquez-Iturbide} {and} \bibinfo{person}{Antonio P{\'e}rez-Carrasco}.} \bibinfo{year}{2016}\natexlab{}.
\newblock \showarticletitle{How to Use the SRec Visualization System in Programming and Algorithm Courses}.
\newblock \bibinfo{journal}{\emph{ACM Inroads}} \bibinfo{volume}{7}, \bibinfo{number}{3} (\bibinfo{year}{2016}), \bibinfo{pages}{42--49}.
\newblock
\href{https://doi.org/10.1145/2948070}{doi:\nolinkurl{10.1145/2948070}}


\bibitem[Vel{\'a}zquez-Iturbide et~al\mbox{.}(2008)]%
        {velazquez2008srec}
\bibfield{author}{\bibinfo{person}{J~{\'A}ngel Vel{\'a}zquez-Iturbide}, \bibinfo{person}{Antonio P{\'e}rez-Carrasco}, {and} \bibinfo{person}{Jaime Urquiza-Fuentes}.} \bibinfo{year}{2008}\natexlab{}.
\newblock \showarticletitle{SRec: An Animation System of Recursion for Algorithm Courses}.
\newblock \bibinfo{journal}{\emph{ACM SIGCSE Bulletin}} \bibinfo{volume}{40}, \bibinfo{number}{3} (\bibinfo{year}{2008}), \bibinfo{pages}{225--229}.
\newblock


\bibitem[von Mayrhauser and Vans(1993)]%
        {von1993code}
\bibfield{author}{\bibinfo{person}{Anneliese von Mayrhauser} {and} \bibinfo{person}{A~Marie Vans}.} \bibinfo{year}{1993}\natexlab{}.
\newblock \showarticletitle{From Code Understanding Needs to Reverse Engineering Tool Capabilities}. In \bibinfo{booktitle}{\emph{Proceedings of 6th International Workshop on Computer-Aided Software Engineering}}. \bibinfo{publisher}{IEEE Comput. Soc. Press}, \bibinfo{address}{Piscataway, NJ, USA}, \bibinfo{pages}{230--239}.
\newblock
\href{https://doi.org/10.1109/case.1993.634824}{doi:\nolinkurl{10.1109/case.1993.634824}}


\bibitem[Ware(2019)]%
        {ware2012information}
\bibfield{author}{\bibinfo{person}{Colin Ware}.} \bibinfo{year}{2019}\natexlab{}.
\newblock \bibinfo{booktitle}{\emph{Information Visualization: Perception for Design}}.
\newblock \bibinfo{publisher}{Morgan Kaufmann}, \bibinfo{address}{San Francisco, CA, USA}.
\newblock


\bibitem[Wong et~al\mbox{.}(2025)]%
        {wong2025spectrum}
\bibfield{author}{\bibinfo{person}{Novia Wong}, \bibinfo{person}{Nai-Yu Cheng}, \bibinfo{person}{Bruna Oewel}, \bibinfo{person}{Katherine~E Genuario}, \bibinfo{person}{SarahElizabeth Stoeckl}, \bibinfo{person}{Stephen~M Schueller}, \bibinfo{person}{Iftekhar Ahmed}, \bibinfo{person}{Andr{\'e} van~der Hoek}, {and} \bibinfo{person}{Madhu Reddy}.} \bibinfo{year}{2025}\natexlab{}.
\newblock \showarticletitle{'It's a Spectrum': Exploring Autonomy, Competence, and Relatedness in Software Development Processes and Tools}. In \bibinfo{booktitle}{\emph{Proceedings of the 2025 CHI Conference on Human Factors in Computing Systems}}. \bibinfo{publisher}{ACM}, \bibinfo{address}{New York, NY, USA}, \bibinfo{pages}{1--19}.
\newblock
\href{https://doi.org/10.1145/3706598.3713250}{doi:\nolinkurl{10.1145/3706598.3713250}}


\bibitem[Wu and Puntambekar(2012)]%
        {wu2012affordances}
\bibfield{author}{\bibinfo{person}{Hsin-Kai Wu} {and} \bibinfo{person}{Sadhana Puntambekar}.} \bibinfo{year}{2012}\natexlab{}.
\newblock \showarticletitle{Pedagogical Affordances of Multiple External Representations in Scientific Processes}.
\newblock \bibinfo{journal}{\emph{Journal of Science Education and Technology}} \bibinfo{volume}{21}, \bibinfo{number}{6} (\bibinfo{year}{2012}), \bibinfo{pages}{754--767}.
\newblock
\href{https://doi.org/10.1007/s10956-011-9363-7}{doi:\nolinkurl{10.1007/s10956-011-9363-7}}


\bibitem[Yates et~al\mbox{.}(2020)]%
        {yates2020characterizing}
\bibfield{author}{\bibinfo{person}{Rebecca Yates}, \bibinfo{person}{Norah Power}, {and} \bibinfo{person}{Jim Buckley}.} \bibinfo{year}{2020}\natexlab{}.
\newblock \showarticletitle{Characterizing the Transfer of Program Comprehension in Onboarding: An Information-Push Perspective}.
\newblock \bibinfo{journal}{\emph{Empirical Software Engineering}}  \bibinfo{volume}{25} (\bibinfo{year}{2020}), \bibinfo{pages}{940--995}.
\newblock
\href{https://doi.org/10.1007/s10664-019-09741-6}{doi:\nolinkurl{10.1007/s10664-019-09741-6}}


\bibitem[Young et~al\mbox{.}(2024)]%
        {young2024productive}
\bibfield{author}{\bibinfo{person}{Jamaal~Rashad Young}, \bibinfo{person}{Danielle Bevan}, {and} \bibinfo{person}{Miriam Sanders}.} \bibinfo{year}{2024}\natexlab{}.
\newblock \showarticletitle{How Productive Is the Productive Struggle? Lessons Learned From a Scoping Review.}
\newblock \bibinfo{journal}{\emph{International Journal of Education in Mathematics, Science and Technology}} \bibinfo{volume}{12}, \bibinfo{number}{2} (\bibinfo{year}{2024}), \bibinfo{pages}{470--495}.
\newblock
\href{https://doi.org/10.46328/ijemst.3364}{doi:\nolinkurl{10.46328/ijemst.3364}}


\bibitem[Zaqoot et~al\mbox{.}(2019)]%
        {zaqoot2019representational}
\bibfield{author}{\bibinfo{person}{Wisam Zaqoot}, \bibinfo{person}{Lih-Bin Oh}, \bibinfo{person}{Lay~Hoon Seah}, \bibinfo{person}{Elizabeth Koh}, \bibinfo{person}{Fang Zhou}, \bibinfo{person}{Wee-Kek Tan}, {and} \bibinfo{person}{Hock-Hai Teo}.} \bibinfo{year}{2019}\natexlab{}.
\newblock \showarticletitle{Representational Fluency in Education: A Literature Review and the Proposal of a New Instrument}. In \bibinfo{booktitle}{\emph{2019 IEEE International Conference on Engineering, Technology and Education (TALE)}}. \bibinfo{publisher}{IEEE}, \bibinfo{address}{Piscataway, NJ, USA}, \bibinfo{pages}{1--7}.
\newblock
\href{https://doi.org/10.1109/tale48000.2019.9225902}{doi:\nolinkurl{10.1109/tale48000.2019.9225902}}


\bibitem[Zingaro et~al\mbox{.}(2018)]%
        {zingaro2018identifying}
\bibfield{author}{\bibinfo{person}{Daniel Zingaro}, \bibinfo{person}{Cynthia Taylor}, \bibinfo{person}{Leo Porter}, \bibinfo{person}{Michael Clancy}, \bibinfo{person}{Cynthia Lee}, \bibinfo{person}{Soohyun Nam~Liao}, {and} \bibinfo{person}{Kevin~C Webb}.} \bibinfo{year}{2018}\natexlab{}.
\newblock \showarticletitle{Identifying Student Difficulties With Basic Data Structures}. In \bibinfo{booktitle}{\emph{Proceedings of the 2018 ACM Conference on International Computing Education Research}}. \bibinfo{publisher}{ACM}, \bibinfo{address}{New York, NY, USA}, \bibinfo{pages}{169--177}.
\newblock
\href{https://doi.org/10.1145/3230977.3231005}{doi:\nolinkurl{10.1145/3230977.3231005}}


\end{thebibliography}
\end{document}